\documentclass[10pt,preprint]{aastex}

\shorttitle{High-energy transients}
\shortauthors{Sun et al.}

\begin{document}
\title{Extra-galactic high-energy transients: event rate densities and luminosity functions}
\author{Hui Sun\altaffilmark{1,2}, Bing Zhang\altaffilmark{1,2,3}, Zhuo Li\altaffilmark{1,2}}

\altaffiltext{1}{Department of Astronomy, School of Physics, Peking University, Beijing 100871, China; hsun$\_$astro@pku.edu.cn; zhang@physics.unlv.edu; zhuo.li@pku.edu.cn}
\altaffiltext{2}{Kavli Institute for Astronomy and Astrophysics, Peking University, Beijing 100871, China;}
\altaffiltext{3}{Department of Physics and Astronomy, University of Nevada, Las Vegas, NV 89154, USA}

\begin{abstract}
Several types of extra-galactic high-energy transients have been discovered, which include high-luminosity and low-luminosity long-duration gamma-ray bursts (GRBs), short-duration GRBs, supernova shock breakouts (SBOs), and tidal disruption events (TDEs) without or with an associated relativistic jet. In this paper, we apply a unified method to systematically study the redshift-dependent event rate densities and the global luminosity functions (ignoring redshift evolution) of these transients. We introduce some empirical formulae for the redshift-dependent event rate densities for different types of transients, and derive the local specific event rate density, which also represents its global luminosity function. Long GRBs have a large enough sample to reveal features in the global luminosity function, which is best characterized as a triple power law. All the other transients are consistent with having a single power law luminosity function. The total event rate density depends on the minimum luminosity, and we obtain the following values in units of ${\rm Gpc^{-3}~yr^{-1}}$: $0.8^{+0.1}_{-0.1}$ for high-luminosity long GRBs above $ 10^{50}~{\rm erg~s^{-1}}$; $164^{+98}_{-65}$ for low-luminosity long GRBs above $5\times 10^{46}~{\rm erg~s^{-1}}$; $1.3^{+0.4}_{-0.3}$, $1.2^{+0.4}_{-0.3}$, and $3.3^{+1.0}_{-0.8}$ above $ 10^{50}~ {\rm erg~s^{-1}}$ for short GRBs with three different merger delay models (Gaussian, log-normal, and power law); $1.9^{+2.4}_{-1.2}\times 10^4$ above $ 10^{44}~{\rm erg~s^{-1}}$ for SBOs, $ 4.8^{+3.2}_{-2.1}\times10^2 $ for normal TDEs above $10^{44}~ {\rm erg~s^{-1}}$; and $0.03^{+0.04}_{-0.02}$ above $ 10^{48}~ {\rm erg~s^{-1}}$ for TDE jets as discovered by {\em Swift}. Intriguingly, the global luminosity functions of different kinds of transients, which cover over 12 orders of magnitude, are consistent with a single power law with an index of -1.6.

\end{abstract}


\section{Introduction}

Extra-galactic high energy transients are intense cosmological transients whose electromagnetic emission peaks in the X-ray or $\gamma$-ray bands. The study of extra-galactic high-energy transients has remained an active field in astrophysics over the years. Wide-field $\gamma$-ray detectors dedicated to study $\gamma$-ray bursts (GRBs) have led to discoveries of other types of high-energy transients, such as supernova shock breakouts (SBOs) and jets from tidal disruption events (TDEs). Upcoming wide field X-ray telescopes (e.g. Einstein Probe, \citealt{Yuan15}) are expected to significantly enlarge the sample of the known high-energy transients, and probably discover new types. 

GRBs are the main extra-galactic $\gamma$-ray transients. Their durations, usually described by $T_{90}$, range from milliseconds to thousands of seconds. Thanks to the extensive observations led by a list of $\gamma$-ray telescopes, such as {\em BATSE}, {\em HETE-II}, {\em INTEGRAL}, {\em Swift} and {\em Fermi}, our understanding of GRBs has been greatly advanced. Two main types based on their durations are short GRBs (or SGRBs) with $T_{90} < 2$ s, and long GRBs (or LGRBs) with $T_{90} > 2$ s \citep{Kou93}. Observations show that these two types of GRBs have distinct physical origins\footnote{The duration classification sometimes leads to false identification of the physical category of GRBs, see \cite{Zhang09} for a detailed discussion.}: massive star core collapses for LGRBs vs. compact star mergers for SGRBs (see \citealt{KZ15} for a recent review). Afterglow observations led to measurements of the redshifts of GRBs, allowing one to access the energetics of these events. Most LGRBs are found to have a typical isotropic luminosity (1-$10^4$ keV) in the range of $10^{51}\sim10^{53}~{\rm erg~s^{-1}}$, which are called high-luminosity long GRBs (HL-LGRBs). A small fraction of the observed LGRBs, on the other hand, have been detected with peak luminosities less than $ 10^{49}~{\rm erg~s^{-1}}$. Most of these events have distinct observational properties, such as long duration, smooth, single-pulse lightcurves, and are usually referred to as low-luminosity long GRBs (LL-LGRBs). It has also been shown that LL-LGRBs have a much higher event rate density than HL-LGRBs \citep{Sod06}. More importantly, they are found to form a distinct component in the GRB luminosity function \citep{Liang07,Vir09}, suggesting that they have a distinct physical origin. Recent studies \citep[e.g.][]{Cam06,Wax07,wangxy07,Bro11,NS12} suggested that at least some LL-LGRBs may be related to breakouts of trans-relativistic shocks from exploding massive stars.

Lacking a sensitive wide-field X-ray camera, the study of X-ray transients is only in its babyhood. Nonetheless, a few types of extra-galactic X-ray transients have been discovered. TDEs, bright X-ray/UV flares generated when super-massive black holes tidally disrupt stars, have been discovered from the archival X-ray survey data of the missions such as {\em ROSAT}, {\em Chandra}, and {\em XMM-Newton} \citep[e.g.][]{Bad96,KG99}. These transients typically last for months to years, much longer than the duration of GRBs. The discovery of Sw J1644+57 \citep{Bur11} by the {\em Swift} satellite suggested that some TDEs can have super-Eddington luminosities, which point towards a relativistic jet associated with the TDE event. The discovery of a thermal component associated with the prompt emission of the LL-LGRB 060218 led to the suggestion that the signal may be related to an SBO. A serendipitous discovery of an X-ray outburst (XRO) 080109 associated with a nearby supernova SN 2008D \citep{Sod08} suggested that there are indeed high-energy transients (even though with a much lower luminosity than LL-LGRBs) associated with SBOs. This discovery established SBOs as a new type of extra-galactic high-energy transient.

There are several important questions regarding these transients: What are the event rate densities of them (i.e. how often do they occur per unit volume)? How do these event rate densities depend on redshift? What are the luminosity functions of these transients? Do the luminosity functions evolve with redshift? Addressing these questions are essential to understand the progenitor systems of these transients and their cosmological evolution. A cross comparison among different transients may also shed light into possible common underlying physics behind these apparently different events.

In the literature, some studies have been carried out to address these questions for individual transients (e.g. \citealt{Liang07,Vir09,Wan10} for both HL-LGRBs and LL-LGRBs; \citealt{Vir11,Wan14} for SGRBs; \citealt{Sod08} for SBOs; \citealt{Esq08} for normal TDEs; and \citealt{Bur11} for jetted TDEs). Due to the small sample size of some types of these transients, the estimates of their event rate densities are usually subject to large uncertainties. The total event rate density of a particular transient depends on the minimum luminosity and the shape of the luminosity function, which is usually not well constrained. Also, the detectors' sensitivity, search algorithms, as well as instrumental selection effects all introduce additional uncertainties to the problem. The calculations of the intrinsic event rate density rely on the sensitivity, field of view, and working period of the detectors. Since these transients have been detected using very different detectors with different sets of parameters, special care needs to be taken in order to obtain robust results.

In order to study the evolution of luminosity functions, one needs a large enough sample that cover a wide redshift range, with each redshift bin having enough events to construct a statistically meaningful luminosity function in the redshift bin. The X-ray transients we are studying mainly reside in the nearby universe, so that their redshift evolution, if any, cannot be investigated. We therefore mainly focus on the luminosity function evolution of GRBs. In the literature, there has been intense discussion about the evolution effect of the luminosity function of long GRBs \citep[e.g.][]{LR02,Yone04,KL06,Sal09,Sal12,Petro15,Yu15,Pescalli15}. Using either a flux-limited sample or a non-parametric method \citep{EP92} to account for the truncation effect, these studies suggested that the LGRB data are consistent with having a redshift-evolving luminosity function. Some papers \citep[e.g.][]{LR02,Petro15,Yu15,Pescalli15} suggested that the data are consistent with the hypothesis that the luminosity function is a broken power law with a universal shape (i.e. the power law indices before and after the luminosity break), but the break itself has a redshift evolution in the form of $L_b \propto (1+z)^k$, with $k \sim 2.3$. 

In this paper, we apply a unified method to systematically study the redshift-dependent event rate densities and the {\em global luminosity functions} (i.e. luminosity functions derived ignoring possible redshift evolution) of several known extra-galactic high-energy transients. For GRBs, thanks to their large sample size, we also investigate their luminosity function evolution. Compared with previous studies, our analysis has a larger sample for most transients (especially for LL-LGRBs, and TDEs), and more interestingly, we will derive the global event rate density distributions of {\em all transients} for the first time. In Section 2, we introduce the general methodology of calculating the event rate density and the luminosity function of any type of transient. We then introduce redshift distributions of various transients in Section 3, especially the new empirical models for short GRBs and TDEs. In section 4, we describe our data of all extra-galactic high energy transients. The results for individual events are presented in Section 5.1 (LGRBs), 5.2 (SGRBs) 5.3 (SBOs), and 5.4 (TDEs), respectively. In 5.5, we present the global distribution of the luminosity functions of all the transients. Conclusions are drawn in Section 6 with some discussion. Throughout the paper, the concordance cosmological parameters presented by the Planck Collaboration, i.e. $ H_{0}=67.8{\rm \,km\,s^{-1}\,Mpc^{-1}} $, $ \Omega_{m}=0.308$, $\Omega_{\Lambda}=0.692$, are adopted \citep{Pla15}.

\section{Methodology}
\subsection{Global Luminosity Function}

Quantifying the redshift-dependent event rate density and luminosity function of a certain type of transient is a challenging task. The observed events and their redshift and luminosity distributions are the results of the convolution of both (likely redshift-dependent) luminosity function and intrinsic redshift distribution of the events, with the proper correction of the instrumental sensitivity threshold, field of view, and operational time. All these complications may be disentangled with a large enough observational sample. However, for most transients discussed in this paper (except GRBs), the number of observational sample is too small to perform such a task. In order to cross-comparing various types of transients, in this paper, we first ignore the possible redshift-evolution of the luminosity functions of all the transients and use the data to construct luminosity functions of each type of transients. This allows us to separate luminosity function and redshift distribution $f(z)$. In principle, the luminosity function could be redshift-dependent. For example, for long GRBs for which we have collected a large enough sample, evidence of such an evolution effect has been collected. As a result, the luminosity function we construct in this paper only carries the meaning in the ``global'' sense. We hereby define all the luminosity functions constructed without considering redshift evolution as {\em global luminosity functions} (GLF). In order to use the data to construct the GLF, we also assume that the events with a same luminosity share the same other properties (e.g. spectral properties and detector parameters). This makes the observed events good indicators of the underlying general population. Also, since there is a wide range of the spectral peak energy ($ E_{\rm peak} $) distribution for different transients studied in this paper, we try as much as possible to apply the $k$-corrected bolometric luminosity ($\rm 1-10^4\,keV $) using the measured spectral properties of the transients (Eq.(\ref{eq:k1})). The only exceptions are the TDEs detected by {\em ROSAT}, {\em Chandra}, and {\em XMM-Newton} whose narrow bandpass does not allow a precise inference of the global spectral parameters. For these events, we use a uniform $k$-correction parameter 1.4 (Eq.(\ref{eq:k2})).

For a certain type of transient, we define the local specific event rate density (local event rate density per unit luminosity) as
\begin{equation}
\rho_{\rm 0,L}=\frac{d\rho_{0}}{dL},
\label{eq:dfrho_0L}
\end{equation}
where $L$ denotes bolometric luminosity at the peak time hereafter. For a detector with flux sensitivity $ F_{th} $, field of view $ \Omega $, and operational time $T$, the detected number of events in the luminosity interval from $ L $ to $ (L+dL) $ is 
\begin{equation}
dN=\frac{\Omega T}{4\pi}\rho_{0,L} dL \int_{0}^{z_{\rm max}(L)}\frac{f(z)}{1+z}\frac{dV(z)}{dz}dz = \frac{\Omega T}{4\pi}\rho_{0,L}g(L)dL,
\label{eq:dN1}
\end{equation}
where 
\begin{equation}
g(L)=\int_{0}^{z_{\rm max}(L)}\frac{f(z)}{1+z}\frac{dV(z)}{dz}dz.
\end{equation}
The dimensionless function $f(z)$ describes the redshift-dependent event rate density, i.e.
\begin{equation}
\rho_{\rm L}(z) = \rho_{\rm 0,L} f(z).
\label{eq:fz}
\end{equation}
The redshift-dependent specific comoving volume reads (for the standard $\Lambda$CDM cosmology)
\begin{equation}
\frac{dV(z)}{dz}=\frac{c}{H_0}\frac{4\pi D_L^2}{(1+z)^2[\Omega_M(1+z)^3+\Omega_{\Lambda}]^{1/2}}.
\end{equation}
For a particular $L$, the maximum redshift $ z_{\rm max}(L)$, which defines the maximum volume inside which an event with luminosity $L$ can be detected, can be defined by the sensitivity threshold $F_{\rm th}$ via
\begin{equation}
F_{\rm th} = \frac{L}{4\pi D_{L}^{2} (z_{\rm max})k},
\end{equation}
where $k$ is a correction factor, which corrects the observed flux in the detector's energy band ($e_1, e_2$) to a wide band in the rest frame (e.g. $1-10^4$ keV for GRBs, see Eq.(\ref{eq:kdf})).

Technically, it is easier to evaluate numbers in the logarithmic luminosity bins. Equation (\ref{eq:dN1}) can be also written as
\begin{equation}
d N=\frac{dN}{d \log L}d \log L=\frac{\Omega T }{4\pi}(\ln10)\rho_{\rm 0,L}g(L)Ld(\log L)
\label{eq:dN2}
\end{equation}
Suppose $\Delta N$ events are detected in a finite logarithmic luminosity bin from $\log L$ to $\log L + \Delta (\log L)$, one then has
\begin{equation}
\rho_{\rm 0,L} \simeq \frac{4\pi}{\Omega T}\frac{1}{\ln10}\frac{1}{g(L)}\frac{1}{L}\frac{\Delta N}{\Delta(\log L)}.
\end{equation}

The luminosity function of a certain type of transient can be defined as
\begin{equation}
 N(L) dL \propto \Phi(L)dL,
\end{equation}
with the integration of $\Phi(L)$ normalized to unity, i.e.
\begin{equation}
\int_{L_{\rm min}}^{L_{\rm max}}\Phi(L)dL=1,
\end{equation}
where $L_{\rm min}$ and $L_{\rm max}$ are the minimum and maximum values of the luminosity distribution. One can define the local event rate density above a certain luminosity $L$, i.e.
\begin{equation}
\rho_{0,>L}=\int_{L}^{L_{\rm max}}\rho_{0,L}dL \simeq \sum_{\log L}^{\log L_{\rm max}}\frac{4\pi}{\Omega T}\frac{1}{\ln10}\frac{1}{g(L)}\frac{\Delta N}{\Delta(\log L)} \frac{\Delta L}{L}
\label{eq:rho0'}
\end{equation}
The total local event rate density is therefore
\begin{equation}
\rho_0=\rho_{0,>L_{\rm min}}=\int_{L_{\rm min}}^{L_{\rm max}}\rho_{0,L}dL \simeq \sum_{\log L_{\rm min}}^{\log L_{\rm max}}\frac{4\pi}{\Omega T}\frac{1}{\ln10}\frac{1}{g(L)}\frac{\Delta N}{\Delta(\log L)}\frac{\Delta L}{L}
\label{eq:rho0}
\end{equation}
which depends on $L_{\rm min}$. Observationally $L_{\rm min}$ is not well constrained, and one usually adopts the {\em observed} minimum luminosity, which is the upper limit of the true $L_{\rm min}$. As a result, the derived $\rho_0$ is in principle only the lower limit of the true value. To be specific, throughout the paper, we always specify a minimum luminosity whenever an event rate density is quoted.

With the definition of $\rho_0$ (Eq.(\ref{eq:rho0})), the specific event rate density can be also written as
\begin{equation}
 \rho_{\rm 0,L} = \rho_0 \Phi(L).
\label{eq:rho_0L}
\end{equation}
Within the framework that the luminosity function does not depend on redshift (the approach adopted in this paper), the redshift-dependent event rate density can be written as
\begin{equation}
 \rho(z) = \rho_{0} f(z),
\end{equation}
where $f(z)$ is the redshift evolution function, the form of which depends on the properties of the transients.

The luminosity function of a certain type of transient can be derived by displaying the specific event rate density $\rho_{\rm 0,L} \propto \Phi(L)$ as a function of $L$. By separating the data into different luminosity bins, we use the observed numbers to map the relevant $\rho_{\rm 0,L}$, and then fit the data points by several empirical model forms. The simplest model is a single power law (SPL) form, i.e.
\begin{equation}
\Phi(L) \propto L^{-\alpha}. 
\end{equation}
If the model does not define the data well, we introduce a smoothly connected broken power law (BPL) form
\begin{equation}
\Phi(L) \propto \left[ \left( \frac{L}{L_b}\right)^{\omega\alpha_1}+\left(\frac{L}{L_b}\right)^{\omega\alpha_2}\right] ^{-1/\omega},
\end{equation}
where $\alpha_1$ and $\alpha_2$ are the power law indices before and after the break luminosity $L_b$, and $\omega$ defines the sharpness of the break.
In more complicated cases (e.g. the GLF of long GRBs), one needs another power law segment to fit the data, and we introduce a triple power law (TPL) form, i.e.
\begin{equation}
\Phi(L) \propto \left\lbrace  \left[ \left(   \left(\frac{L_{b,1}}{L_{b,2}}\right) ^{\omega_1\alpha_2}+\left(\frac{L_{b,1}}{L_{b,2}}\right)^{\omega_1\alpha_3} \right) ^{-1/\omega_1} \left( \frac{L}{L_{b,1}}\right) ^{-\alpha_1}  \right] ^{-\omega_2}+ \left[ \left( \left(\frac{L}{L_{b,2}}\right)^{\omega_1\alpha_2}+\left(\frac{L}{L_{b,2}}\right)^{\omega_1\alpha_3}\right)^{-1/\omega_1} \right]^{-\omega_2}  \right\rbrace ^{-1/ \omega_2},
\end{equation}
where $\alpha_1$, $\alpha_2$, and $\alpha_3$ are the power law indices for three segments, $L_{b,1}$ and $L_{b,2}$ are the two break luminosities, and $\omega_1$ and $\omega_2$ are the sharpness parameters at the two breaks.

\subsection{Luminosity function evolution of GRBs}\label{sec:evolution}
GRBs have a large enough sample to study the redshift evolution effect. We investigate the evolution effect for LGRBs and SGRBs separately. For each class, we first separate the observed GRBs into several redshift bins, and then apply our method to map the corresponding {\em local specific event rate density} using the GRBs in that specific redshift bin only\footnote{There are some overlaps in adjacent redshift bins, since we want to include more GRBs in each bin to reach a better constraint of the corresponding luminosity function.}. Practically, for a redshift bin $(z_1, z_2)$ around a certain redshift $z$, we change the integration limits in Eq.(2) to $z_1$ and min$(z_2, z_{\rm max}(L))$, respectively, and repeat the procedure laid out in Section 2.1. The derived local specific event rate density and event rate density are expressed as $\rho_{\rm 0,L}^{z}$ and $\rho_{\rm 0,>L_m}^{z}$, respectively, denoting that they are derived in the redshift bin around $z$. Notice that $\rho_{\rm 0,L}^{z}$ still denotes the local value. 
By applying a proper correction with te redshift-evolution function $f(z)$ (see details in Section 3), one can obtain the luminosity function in the redshift bin $(z_1 < z < z_2)$. 
If GRB luminosity function does not evolve with redshift, then the results derived from different $z$ bins should remain the same. To minimize the truncation effect by the flux limits of the detectors, we also use a sub-sample with a higher threshold (with peak photon flux larger than 1.8$\rm~ph~s^{-1}~cm^{-2}$) in the derivations, and compare the results with two thresholds.

Alternatively, we also repeat the analysis under the same assumption adopted in some recent papers \citep[e.g.][]{Petro15,Yu15,Pescalli15}, i.e. fix the shape of the luminosity function and assume that $L_b$ evolves as a power law with $k\sim2.3$. We can then map the luminosity function from the data by correcting the luminosity $L$ at $z$ to the ``local'' value, i.e. $L_0=L/(1+z)^{2.3}$. Applying the method in Section 2.1 using $L_0$ instead of $L$ would lead to the ``local'' luminosity function.

\subsection{Correction factors}

To perform our analysis, the redshift of an event is needed. For GRBs, not all events have redshift measurements. In order to properly account for their event rate density, one needs to correct the derived values based on the $z$-known sample by the ratio between the total number and the $z$-known event number. For HL-LGRBs and SGRBs, this correction factor is approximately 3, and we adopt it in our derivations.

The detected rates also depend on the detector's spectral window. For GRBs, {\em BATSE} observations suggested that the short-to-long ratio is about 1:3 \citep{paciesas99}, whereas {\em Swift} BAT, which has a softer bandpass, only gets a 1:10 ratio \citep{sakamoto11,qin13}. Since {\em Swift} can essentially detect all {\em BATSE} LGRBs, one needs to correct for the {\em Swift}-detected SGRBs by another factor of $\sim 3$ for the inferred SGRB event rate density. This factor has been taken into account in our derivations.

For the other types of transients, the samples are not large enough to access whether we have missed some events due to the imperfect instrumental spectral window or the lack of redshift measurements. We therefore derive the event rate density using the observed events only.

\subsection{Instrumental parameters}

The three important instrumental parameters that are relevant to our derivations are the operation time ($T$), the field of view ($\Omega$), and the flux sensitivity threshold ($F_{\rm th}$). Table \ref{tab:instruments} lists the three parameters of the high-energy detectors used to study various high-energy transients discussed in this paper. While the first two parameters are straightforwardly defined, the definitions of sensitivity thresholds are non-trivial. This is particularly true for wide field triggering detectors such as {\em Swift} BAT, with which most of the GRBs and jetted TDEs were detected (e.g. \citealt{Lie14} for a detailed description of {\em Swift} BAT trigger algorithm). In this paper, we adopt an approximate threshold for each detector. For {\em Swift} detected events, we also adopt slightly different values for different types of events. 

Most GRBs were detected through the rate trigger algorithm by BAT, whereas some low-flux events, such as the LL-LGRB 060218 were detected through an image trigger algorithm. The image trigger is an additional trigger algorithm to accumulate photons from a source in a comparably longer time to look for transients that are not bright enough to make a rate trigger. 

For the rate-trigger GRB events detected by Swift/BAT, we adopt slightly different values for different sub-types. In order to have the redshift of a GRB measured, the burst usually needs to have a bright enough optical afterglow. On average, the $z$-known GRBs are brighter and therefore have a higher flux threshold than the standard rate trigger flux threshold. Based on the lowest value of the flux distributions of our sub-sample, we adopt  $F_{\rm th} =  3\times10^{-8}~{\rm erg\,cm^{-2}\,s^{-1}} $ for rate-triggered LGRBs (including HL-LGRBs and LL-LGRB 080517), and $10^{-7}~{\rm erg\,cm^{-2}\,s^{-1}}$ for SGRBs, respectively. 

Taking HL-LGRBs as an example, we derive the flux threshold based on the lower end of the observed photon flux distribution. We adopt the photon flux 0.3 $\rm ph\,cm^{-2}\,s^{-1}$ as the threshold\footnote{Out of 250 HL-LGRBs in our sample, only two have peak photon flux below 0.3 $\rm ph\,cm^{-2}\,s^{-1}$. One of them (GRB 070419A) was detected through image trigger, and the other (GRB 060123) did not trigger BAT but was detected from the BAT survey data.}. The transformation from photon flux to specific flux (Eq. (\ref{eq:pf})) requires the information of the spectrum. We consider a typical Band function spectrum \citep{Ban93} with $\alpha=-1$. $ \beta=-2.3$, $ E_{\rm peak}=200$ $\rm keV$ at redshift $z=1$. This threshold photon flux is translated to $F_{\rm th} =  3\times10^{-8}~{\rm erg\,cm^{-2}\,s^{-1}}$.

The LL-LGRBs 060218 \& 100316D and the two {\em Swift}-detected TDE jets were detected through image trigger. In the case of image trigger algorithm, the threshold flux depends on the trigger duration $T_{\rm td}$, with the dependence $F_{\rm th} \propto T_{\rm td}^{-1/2}$. The trigger duration of GRB 060218 was about 80 s\footnote{http://gcn.gsfc.nasa.gov/other/191157.swift}, with a mean flux of $\sim 2.8 \times 10^{-9}~{\rm erg~cm^{-2}~s^{-1}}$ (T. Sakamoto, 2015, private communication). This roughly corresponds to a threshold flux
\begin{equation}
F_{\rm th}=2.5\times10^{-8}T_{\rm td}^{-1/2}~{\rm erg~cm^{-2}~s^{-1}},
\end{equation}
which we adopt to also calculate the threshold flux for other image trigger events. The trigger duration of GRB 100316D is 64 s\footnote{http://gcn.gsfc.nasa.gov/other/416135.swift}. The two jetted TDE events Sw J1644+57 and Sw J2058+05 had a trigger duration of 64s\footnote{http://gcn.gsfc.nasa.gov/other/450258.swift} \citep{Bur11} and 4d \citep{Cen12}, respectively, and the corresponding $F_{\rm th}$ are used to derive $\rho_{\rm 0,L}$ of jetted TDEs. In fact, Sw J1644+57 was image-triggered four times. The 64s trigger duration was the relevant one at the peak luminosity.

XRO 080109 was serendipitously discovered by {\em Swift} X-Ray Telescope (XRT). We use a count rate of 0.03 counts/s for XRT with which source variability can be detected for a $\sim 1000$ s observation. This corresponds to a flux threshold $F_{\rm th} \sim 10^{-12}~{\rm erg\,cm^{-2}\,s^{-1}}$.

Normal X-ray TDEs were usually discovered from archival survey data of various X-ray missions, such as the {\em ROSAT} PSPC All-Sky Survey (RASS) \citep{Vog99}, the {\em XMM-Newton} Slew Survey Source Catalogue (XMMSL1) \citep{Sax08}, and the {\em Chandra} ACIS archival data. The exposure-time-dependent flux sensitivity thresholds of these three detectors are listed in Table \ref{tab:instruments}. For each TDE event, we consider the real exposure time to determine its $F_{\rm th}$.

\section{Redshift distribution}

The redshift distribution parameter $f(z)$ for each type of transients is essential to infer the local (specific) event rate density (\S2), and different types of transients may have different $f(z)$ functions. In this section, we discuss this function for different types of transients in detail.

\subsection{Long GRBs and supernova shock breakouts}

Long GRBs (both HL and LL) and SBOs are associated with the deaths of massive stars. To first order, their redshift distribution should track the history of star formation. There is evidence that at high-$z$, the GRB rate may exceed what the star formation history (SFH) predicts \citep[e.g.][]{li08,Kistler08,Qin10,Virgili11b,Robertson12}. However, for a wide redshift span, the SFH is a good proxy of the redshift distribution of LGRBs. In this paper, we adopt the rough analytical model of SFH derived by \cite{Yuk08} using the observational data:
\begin{equation}
f_{\rm LGRB/SBO}(z)=\left[ (1+z )^{3.4\eta}+\left( \frac{1+z}{5000}\right) ^{-0.3\eta}+\left( \frac{1+z}{9}\right) ^{-3.5\eta} \right] ^{\frac{1}{\eta}},
\label{eq:sfh}
\end{equation}
where $\eta=-10$. At $z<4$, this function is directly derived from the SFH inferred from the UV and far-IR galaxy data \citep{HB06}, which is independent on the GRB observations. At $z=5-7$, the SFH is enhanced from the galaxy-constrained SFH to compensate the observed GRB excess at high-$z$, which can be explained by the deficiency of the observed low-luminosity star forming galaxies missed in surveys but traced by GRBs. 
We adopt Eq.(\ref{eq:sfh}) to study both LGRBs and SBOs. The latter are only observed in the nearby universe, so that the modification at high-$z$ does not enter the problem.

\subsection{Short GRBs}
Unlike long GRBs, most short GRBs do not directly trace star formation history. Observations suggest that most short GRBs are consistent with having an origin not related to massive star deaths. The leading scenario is mergers of double compact star systems, e.g. two neutron stars (NS-NS) or a neutron star and a black hole (NS-BH) (\citealt{gehrels04,fox05,barthelmy05,fong10}, see \cite{berger14} for a recent review). 

In order to have a merger to occur, a compact star binary system needs to go through a long inspiral phase defined by energy loss of the system through gravitational wave radiation. The redshift distribution of SGRBs therefore needs to account for an additional time delay due to inspiral with respect to the creation of the compact binary system, which itself traces the SFH. The distribution of the merger delay time scale is unfortunately not known. Practically, one assumes some empirical forms of the merger delay time scale distribution models, and apply the data to derive best parameters for the delay model. Three types of merger delay time models have been discussed in the literature (e.g. \citealt{Vir11}; \citealt{Wan14} and references therein): power law (PL) decay model, Gaussian delay model, and log-normal delay model. Current data support either a Gaussian \citep{Vir11} or log-normal \citep{Wan14} delay model, with the (PL) model disfavored (even though not completely ruled out) \citep{Vir11,Wan14}.  Table \ref{tab:DL} lists the three models with the best parameters currently constrained by the SGRB data.

\begin{figure}[!htb]
\centering
\includegraphics[width=3in]{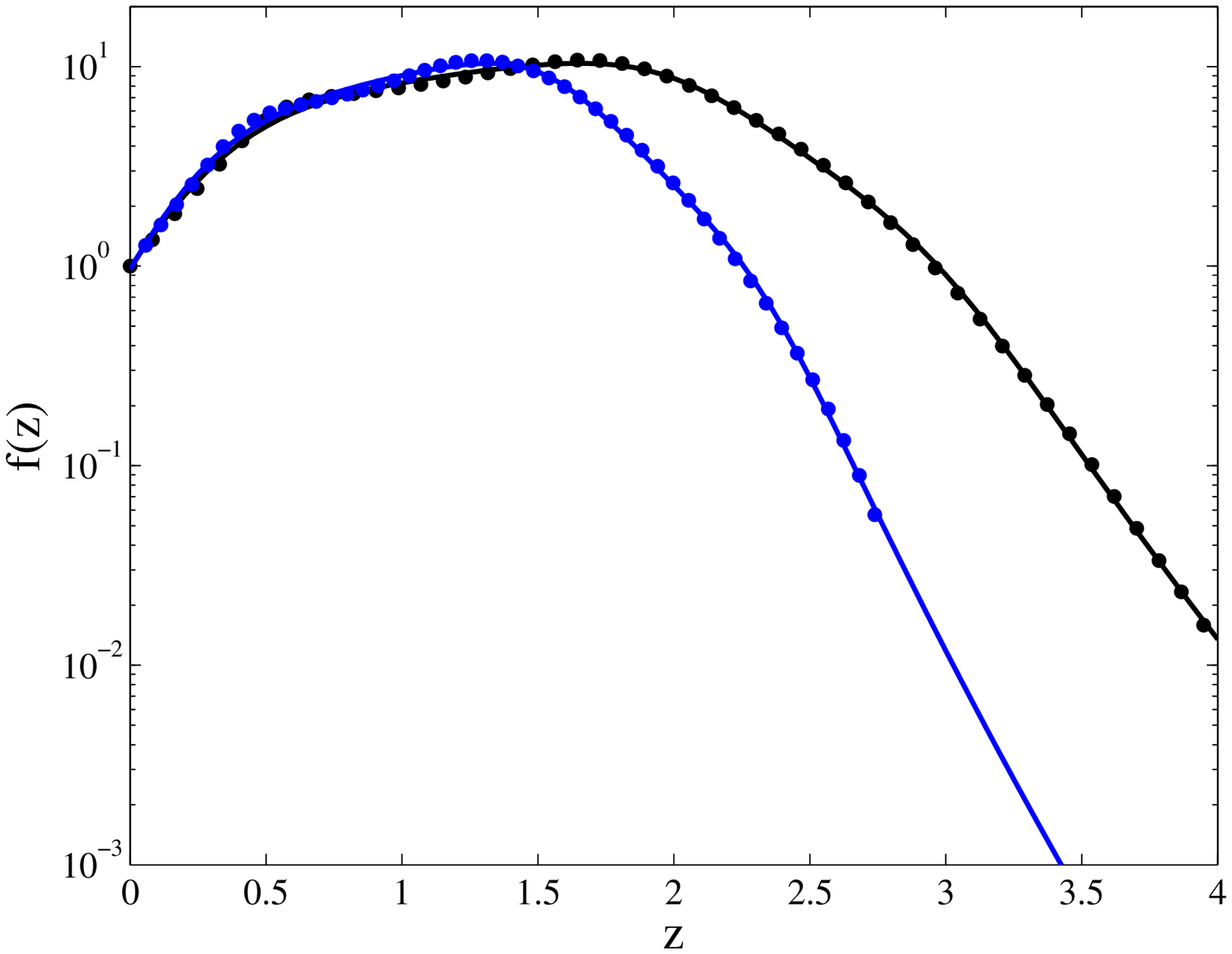} 
\includegraphics[width=3in]{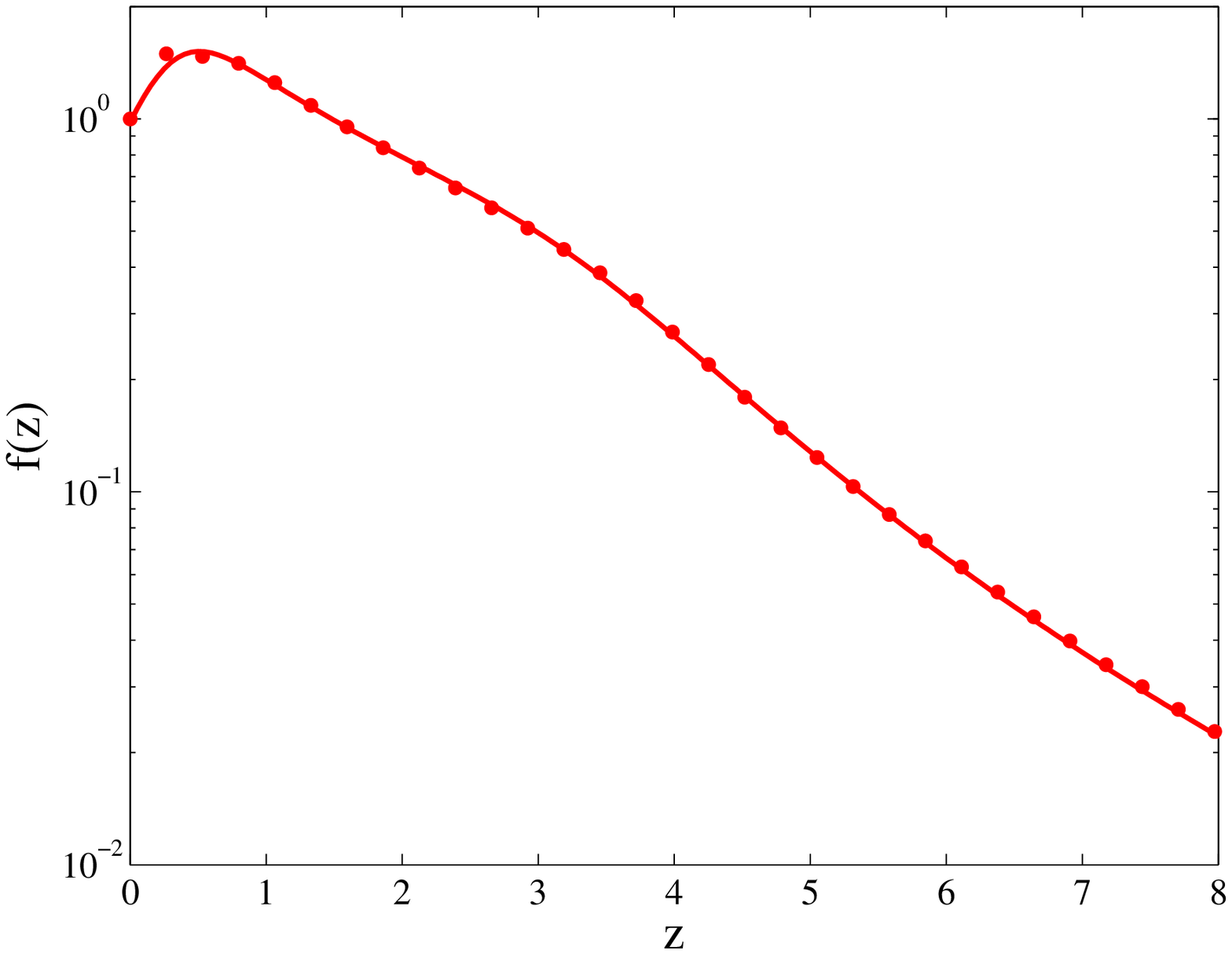} 
\caption{The redshift distribution derived from Monte Carlo simulations for short GRBs considering three delay time models with respect to star formation history: Gaussian (black), log-normal (blue), and power-law (red). For each model, the result is derived from the average of 10000 simulations, each with simulated 10000 systems. Dots are the simulated results, and the curve is the empirical multiple-power-law fits given in Eqs. (\ref{eq:gaussian}), (\ref{eq:ln}) and (\ref{eq:pl}).} 
\label{fig:zsgrb}
\end{figure}

With the consideration of the merger delay time distribution, it is difficult to construct an analytical model for the redshift distribution of SGRBs. Instead, we perform a series of Monte Carlo simulations based on the SFH and merger delay distribution models to construct several redshift distributions that correspond to the three delay models with the best-fit parameters. First, we randomly generate 10,000 compact star binary systems with a redshift distribution tracking the SFH following the model of \cite{Yuk08}. Next, we randomly generate the merger delay time scales of all these systems based on the three merger delay time scale models listed in Table \ref{tab:DL}. For each model, we derive the look back time of SGRBs by subtracting the merger delay time from the formation time, and transfer the lookback time to redshift.  We repeat the process 10,000 times (each with 10,000 events simulated). By averaging the results, we are able to derive the average redshift distribution of the simulated samples. We fit the derived redshift distribution (for all three merger delay models) using multiple-power-law functions and derive an empirical expression of $ f(z) $ for each model. The simulated results with best fit empirical models are shown in Fig.\ref{fig:zsgrb}. The distributions are normalized to unity at the local universe ($z=0$). The empirical formulae of $f(z)$ for the three merger delay models are:

For the Gaussian delay model \citep{Vir11}, one has
\begin{equation}
f_{\rm SGRB}^{\rm G}(z)=\left[ (1+z )^{5.0\eta}+\left( \frac{1+z}{0.17} \right) ^{0.87\eta}+\left( \frac{1+z}{4.12} \right) ^{-8.0\eta} + \left( \frac{1+z}{4.05}\right) ^{-20.5\eta} \right] ^{\frac{1}{\eta}},
\label{eq:gaussian}
\end{equation}
with $ \eta=-2 $, which is roughly a broken power law with redshift breaks at $z_1=0.45$, $z_2=2.0$, $z_3=3.0$.

For the log-normal delay model \citep{Wan14}, one has
\begin{equation}
f_{\rm SGRB}^{\rm LN}(z)=\left[ (1+z )^{5.7\eta}+\left( \frac{1+z}{0.36}\right) ^{1.3\eta}+\left( \frac{1+z}{3.3}\right) ^{-9.5\eta} + \left( \frac{1+z}{3.3}\right) ^{-24.5\eta} \right] ^{\frac{1}{\eta}},
\label{eq:ln}
\end{equation}
with $ \eta=-2 $, which is roughly a broken power law with redshift breaks at $z_1=0.35$, $z_2=1.5$, $z_3=2.3$.

For the power law model \citep{Wan14}, one has
\begin{equation}
f_{\rm SGRB}^{\rm PL}(z)=\left[ (1+z )^{1.9\eta}+\left( \frac{1+z}{2.5}\right) ^{-1.2\eta}+\left( \frac{1+z}{3.8}\right) ^{-4.4\eta} + \left( \frac{1+z}{7.7}\right) ^{-11\eta} \right] ^{\frac{1}{\eta}},
\label{eq:pl}
\end{equation}
with $ \eta=-2.6$. This model has a wider redshift distribution compared to the first two model (due to the wide range of the merger delay time). It is roughly a broken power law with redshift breaks at $z_{1}=0.42$, $z_{2}=3.4$, $z_{3}=11.3$. The SGRB data do not favor this model \citep{Vir11}, even though it is not completely ruled out.

\subsection{Tidal disruption events}

\begin{figure}[!htb]
\centering
\includegraphics[width=3in]{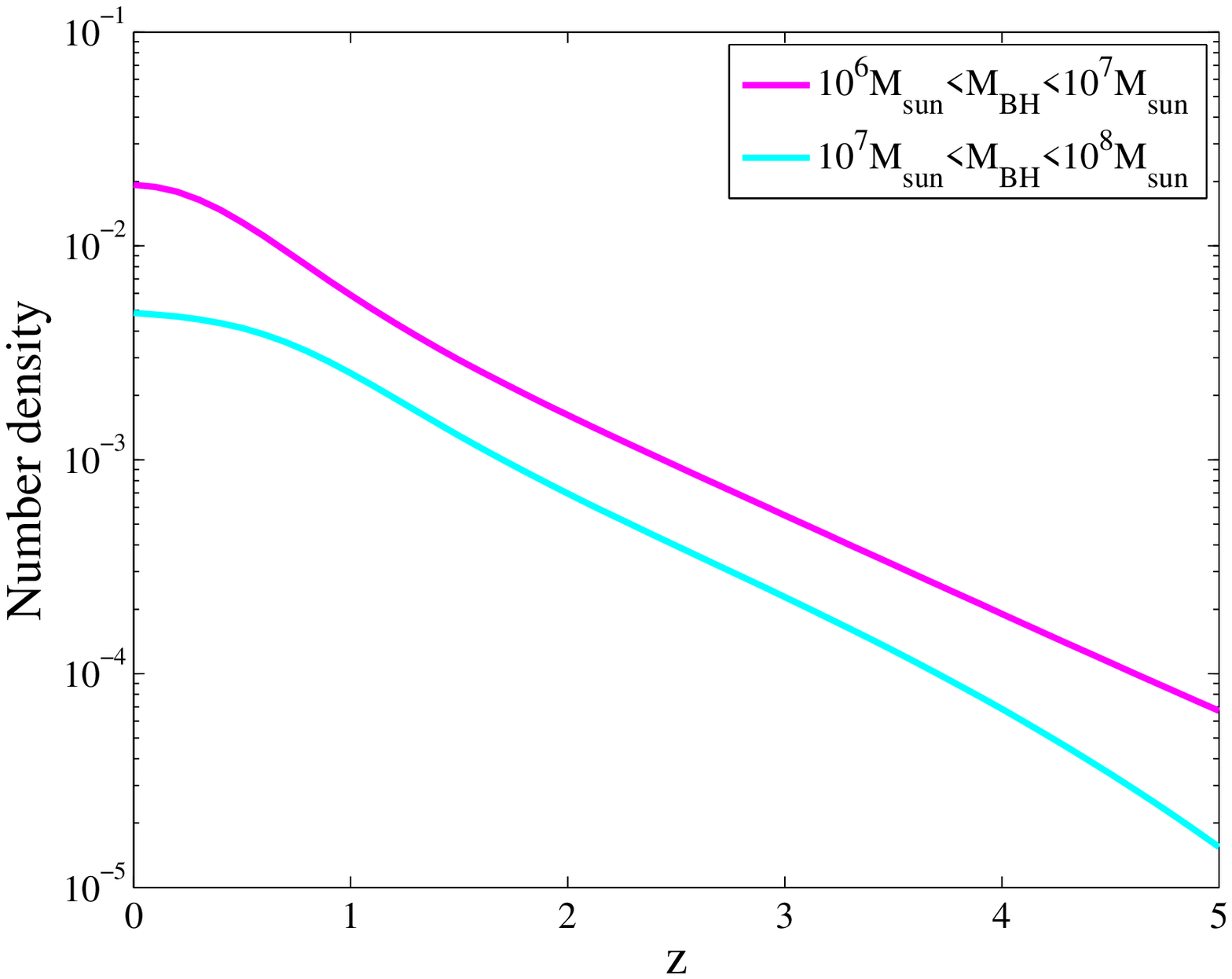}
\includegraphics[width=2.75in]{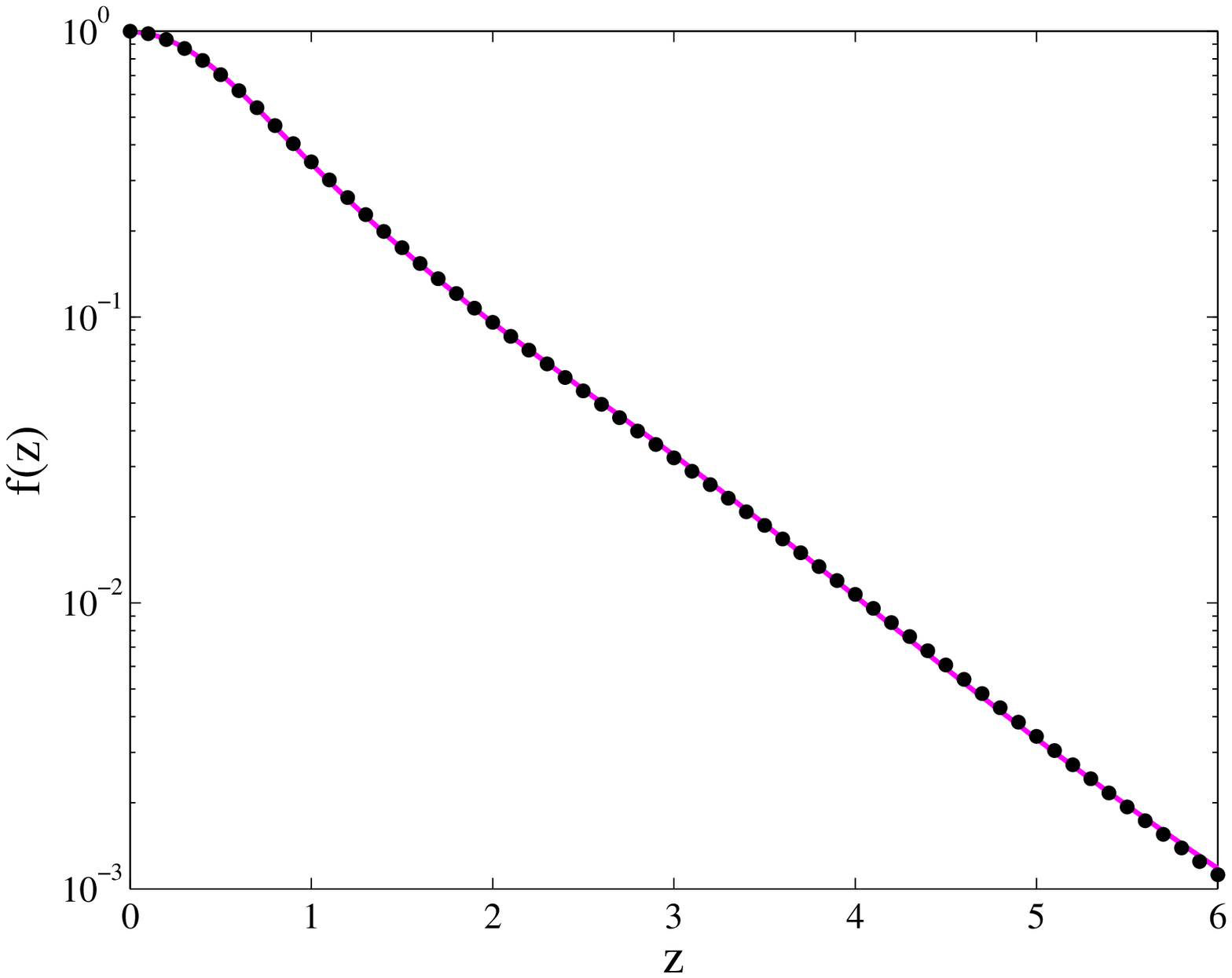}
\caption{Left panel: Redshift-dependent number density of supermassive black holes with masses in the range of $ 10^6-10^7 M_\odot $ (magenta) and $10^7 - 10^8~M_{\odot} $ (cyan) derived from the results of \cite{Sha13}. Right panel: Normalized redshift distribution of TDEs through simulation (dots) and the empirical fit (Eq.(\ref{eq:tde})). \label{fig:ztde}}
\end{figure}

The event rate density of TDEs depends both on the number density of supermassive black holes (SMBHs) and the event rate of TDEs per galaxy. Since the TDE rate of a particular SMBH only depends on the properties of the galaxy itself (e.g. stellar density near the SMBH and the mass of SMBH), on average, it may be reasonable to assume that there is no redshift evolution of the event rate per galaxy. As a result, the $f_{\rm TDE}(z)$ parameter of TDEs is mostly determined by the evolution of the number density of SMBHs as a function of redshift \citep[e.g.][]{Donnarumma15}). \cite{Sha13} constructed the mass density distribution models for SMBHs and AGNs by considering their growth rate and radiation efficiency. We apply their model to derive $f_{\rm TDE}(z)$ for TDEs. TDEs can happen only when the tidal disruption radius is larger than the event horizon of SMBHs, which gives an upper limit of the mass of SMBHs for TDEs:
\begin{equation}
M_{\rm BH}\leq 1.6\times 10^8 \left(\frac{M_{\ast}}{M_{\odot}} \right)\left(\frac{R_{\ast}}{R_{\odot}} \right) ^{3/2}, 
\end{equation}
where $ M_{\ast} $ and $ R_{\ast} $ are the mass and radius of the star that is disrupted by the SMBH, both normalized to the solar values. We therefore exclude SMBHs with mass exceeding $10^8 M_\odot$. In the left panel of Fig.\ref{fig:ztde}, we present the numerical fits to the number density redshift evolution of SMBHs with two mass ranges ($10^6-10^7 M_\odot$ and $10^7-10^8 M_\odot$). Assuming a contant TDE rate per galaxy, in the right panel of Fig.\ref{fig:ztde}, we present the normalized TDE redshift distribution $f_{\rm TDE}(z)$ derived from the numetical data based on the model of \cite{Sha13}. The best-fit emperical model reads
\begin{equation}
f_{\rm TDE}(z)=\left[ (1+z )^{0.2\eta}+\left( \frac{1+z}{1.43}\right) ^{-3.2\eta}+\left( \frac{1+z}{2.66}\right) ^{-7.0\eta} \right] ^{\frac{1}{\eta}},
\label{eq:tde}
\end{equation}
with $ \eta=-2 $. One can see that $f_{\rm TDE}(z)$ continues to decrease with redshift, reaching 2/5 at $z \sim 1$, and $\sim 10^{-3}$ at $z \sim 6$. 

\section{Data}
\subsection{Gamma-ray bursts}
Our HL-GRB sample is only limited to {\em Swift} GRBs. This is because it is a uniform sample whose size is large enough to derive a well-constrained GLF. We collect all the $z$-known {\em Swift} GRBs before May 6, 2014 (250 HL-LGRBs and 20 SGRBs). This sample consists of more GRBs than previous work by introducing a lower flux threshold, which allow us to better study the features near the low luminosity end. The data are downloaded from the {\em Swift} archival table available at http://swift.gsfc.nasa.gov/archive/grb$\_$table/ \citep{Sak08,sakamoto11}. For all the bursts, the 1-s peak photon flux and photon index are recorded. For HL-LGRBs, the 1-s peak photon flux is directly derived from their 1-s peak luminosity. For SGRBs, since their durations are typically shorter than one second, we apply photon count rate with a 64-ms resolution to derive the 64-ms peak luminosity. We calculate their 64-ms peak photon flux based on the ratio of the two peak count rates with different temporal resolutions ($C_{\rm p,64ms}$ and $C_{\rm p,1s}$), i.e. $P_{64} = P_1 (C_{\rm p,64ms}/C_{\rm p,1s})$. The 64-ms light curves are from the {\em Swift} Burst Ground-Analysis Information page (http://gcn.gsfc.nasa.gov/swift$\_$gnd$\_$ana.html), and the 1-s light curves are obtained through re-binning. To ensure the correct match at the peak, the regrouping is such that the time interval at the peak time (64-ms resolution) matches the one for the 1-s peak photon flux provided at the GCN Circular archive. Only a handful of LL-LGRBs were detected so far. Table \ref{tab:GRB} collects the information of six LL-LGRBs studied in this paper, which were triggered not only by {\em Swift} (GRBs 060218, 080517 and 100316D), but also by other instruments as well: GRB 980425 by {\em CRGO}/BATSE, XRF 020903 by {\em HETE-II}\footnote{The acronym ``XRF'' stands for ``X-ray flashes''. They are softer version of GRBs. Observations show that XRFs and GRBs seem to form a continuum in the observational and theoretical parameter spaces \citep{Sak04,Zha04,Ber06}. In fact, GRB 060218 can be also called an XRF. We adopt the names of these events based on the convention adopted in their discovery papers.}, and GRB 031203 by {\em INTEGRAL}. The peak photon fluxes of pre-Swift LL-LGRBs are adopted from GCN archives or \cite{Sak04}. The data of {\em Swift} LL-LGRBs are also taken from the {\em Swift} Table.

The time integrated spectral information is taken from the literature (references provided in Table \ref{tab:GRB}), described by either a single power law (PL) with photon index $ \Gamma  (N(E)\propto E^{-\Gamma} $), a Band function characterized by peak energy $E_{\rm peak}$ and two photon spectral indices $\alpha $ and $\beta $  \citep{Ban93}, or a power law function with an exponential cutoff (CPL) fit, i.e. $N(E)\propto E^{\alpha} \exp(-E/E_c) $. For the latter two models, an $ E_{\rm peak}$ can be derived from the peak in the $\rm \nu F_{\nu}$ spectrum. For single PL fits to most BAT spectra (due to the narrowness of the BAT band), it is believed that the intrinsic spectrum still has a peak energy. With BAT GRBs jointly detected by other wide-band detectors such as {\em Konus}/Wind and {\em Fermi}/GBM, it was found that there exists a rough correlation between the BAT-band photon index $ \Gamma $ and $ E_{\rm peak}$, if $ E_{\rm peak} $ is not much beyond the BAT energy band \citep{Zhang07,Sak09,Vir12}. The latest best fit reads \citep{Vir12}
\begin{equation}
\log(E_{\rm peak})=(4.34\pm 0.475)-(1.32\pm 0.129)\Gamma^{\rm BAT}
\label{eq:Ep-Gamma}
\end{equation}
 with a large scatter, where $ \Gamma^{\rm BAT} $ is photon index (positive value) defined in the BAT band.  We apply this scaling to estimate $E_{\rm peak}$ for those GRBs whose $E_{\rm peak}$ is not directly measured. For a consistency check, we have also adopted $E_{\rm peak}$ values derived by \cite{Butler07} for a sub-sample of GRBs (for which $E_{\rm peak}$ is available from that method). By repeating the calculations, we found that the derived LF using the \cite{Butler07} method is similar to the LF derived using our method. 

For a GRB with peak photon flux $ P_{p} $, the peak flux can be calculated through
\begin{equation}
F_{p}=\frac{P_{p}\int_{e_{1}}^{e_{2}}EN(E)dE}{\int_{e_{1}}^{e_{2}}N(E)dE},
\label{eq:pf}
\end{equation}
where $N(E)$ is the photon spectrum of a GRB, which is in the form of the standard Band function \citep{Ban93}
\begin{equation}
N(E) = A \left\{
 \begin{array}{ll}
  \left( \frac{E}{\rm 100 keV}\right)^\alpha \exp(-\frac{E}{E_0}), & E < (\alpha-\beta)E_0, \nonumber \\
  \left[\frac{(\alpha-\beta) E_0}{\rm 100 keV}\right]^{\alpha-\beta} \exp(\beta-\alpha) \left(\frac{E}{\rm 100 keV}\right)^{\beta}, & E \geq (\alpha-\beta) E_0.
 \end{array}
\right.
\end{equation}
Here the integration limits $ (e_{1},e_{2}) $ enclose the detector's energy window (e.g. 15-150 keV for {\em Swift} BAT). For short GRBs, we use the $ E_{\rm peak} $ data of 13 SGRBs derived by \cite{Lv15}. For other 7 SGRBs whose $E_{\rm peak}$ was not directly measured, we estimate $ E_{\rm peak} $ using Eq.(\ref{eq:Ep-Gamma}). For those GRBs whose Band function parameters are not directly measured, we adopt typical values as $ \alpha=-1 $ and $ \beta=-2.3 $ for LGRBs and $ \alpha=-0.5 $ and $ \beta=-2.3 $ for SGRBs.

In order to derive the bolometric luminosity ($1-10^4$ keV in the cosmological rest frame) from the observed peak flux, we perform a $k$-correction 
\begin{equation}
L_{p,\rm bol}=4 \pi D_{L}^{2}F_{p}\cdot k
\label{eq:k1}
\end{equation}
where $ D_{L} $ is the luminosity distance. The $k$-correction parameter can be expressed as
\begin{equation}
k=\frac{\int_{1/(1+z)}^{10^{4}/(1+z)} EN(E)dE}{\int_{e_{1}}^{e_2} EN(E)dE}.
\label{eq:kdf}
\end{equation}

\subsection{Shock breakouts}

Massive stars end their lives in catastrophic core collapses when they run out of fuel in the center \citep{WW86}. As a massive star undergos core collapse, an outgoing shock surges through the star. When the optical depth of photons trapped in the shock becomes unity, a SBO occurs, which provides the first electromagnetic emission from a supernova event.  Before the shock breaks out the star, only neutrinos and gravitational waves can escape. At the moment of breakout, a short, bright flash is expected, which peaks in ultravoilet or X-rays depending on how compact the star is 
\citep{Col75,KC78,NS10}. The SBO signal therefore carries direct information about the very early stage of core collapse and provides direct constraints on the type of progenitor. Since there is no electromagnetic precursor to alert such an event, detecting an SBO is challenging. In the X-ray and soft $\gamma$-ray regime for which our paper focuses on, there are only two confirmed SBOs detected so far. One is GRB 060218/SN 2006aj association system, which shows an X-ray thermal component with a temperature of $\sim 0.17$keV in a very long duration ($T_{90}=2100$s), soft GRB with a smooth lightcurve \citep{Cam06}. The other is X-Ray Outburst (XRO) 080109/SN 2008D association system, which was serendipitously detected by {\em Swift}/XRT on 2008 January 9 \citep{Sod08}. Since no $\gamma$-ray counterpart was detected even though this outburst was in the field of view of BAT before and during the burst, a GRB connection was ruled out. These two observations have offered a great opportunity to study the detailed properties about the progenitors. The fact that several other LL-LGRBs seem to share similar properties to GRB 060218 makes some authors suggest that all LL-LGRBs may be associated with SBOs \citep[e.g.][]{wangxy07,NS12}.

We use the two confirmed SBO events (with very different luminosities) to estimate their event rate densities. The data of GRB 060218/SN 2006aj are already included in the Table of LL-LGRBs. The data of XRO 080109/SN 2008D are collected from \cite{Sod08}. Its X-ray spectrum was fit by a power-law model with a photon index of 2.61 in the band of $\rm 0.3-10$ keV, which was used for $k$-correction.

\subsection{Tidal disruption events}

Stellar tidal disruption by a supermassive black hole has been theoretically predicted by \cite{Ree88}. When a star approaches a supermassive black hole, a tidal disruption event (TDE) would occur if the tidal force becomes larger than the self-gravity of the star and if the radius when this happens is outside the black hole event horizon. Part of the disrupted debris falls into the black hole from an accretion disk, giving rise to a bright flare in UV or X-ray band, which lasts for several months to one year. The first TDE was discovered in a quiescent galaxy NGC 5905 during the RASS survey \citep{Bad96,KB99}, which showed the characteristic luminosity decay law $ L\propto t^{-{5}/{3}} $ expected for fall-back accretion. Later, several more TDEs have been detected from RASS, namely RX J1242, RX J1624, RX J1420 (\citealt{KG99}; \citealt{Gru99}; \citealt{Gre00}). In the recent decade, more TDE candidates have been discovered by {\em XMM-Newton} and {\em Chandra}, mostly by comparising survey catalogs with the archival data (\citealt{Esq07}; \citealt{Esq08}; \citealt{Mak10}; \citealt{Sax12}; \citealt{Mak13}), as well as by serendipitous detections \citep{Lin11}. Another {\em ROSAT} source was identified as a TDE candidate as it disappeared in the subsequent observations with {\em XMM-Newton} and {\em Chandra} \citep{Cap09}. Right now about a dozen X-ray TDE candidates have been discovered. All these TDEs have large amplitudes and soft X-ray spectra, whose host galaxies show no sign of AGN activity \citep{Kom12}. Their observed maximum luminosities range from $\rm ~10^{42}-10^{45}\,erg\,s^{-1} $. 

Recently two special TDEs were detected by {\em Swift}. These two TDEs, i.e. Swift J1644+57 \citep{Blo11,Bur11} and Swift J2058.4+0516 \citep{Cen12}, showed some distinct features. The peak luminosity of Sw J1644+57 was around $\rm 10^{48}\,erg\,s^{-1} $, and the event was followed by a radio counterpart \citep{zauderer11}. Due to the super-Eddington nature of the events, these TDEs have been interpreted as relativistic jets launched from the central black hole (\citealt{Blo11}; \citealt{Bur11}). It has been claimed that there is a low probability that normal TDEs also host a jet similar to Swift J1644+57 \citep{Van13}. However, it remains unknown whether two types of TDEs are indeed intrinsically different from others, and if so, what could be the main reason to make the difference. By measuring the spin parameter of the central black holes of the two {\em Swift} TDEs within the theoretical framework of the Blandford-Znajek mechanism, \cite{LZ11} found that both black holes carry a moderately high spin. They then suggested that black hole spin may be the key factor to make the dichotomy of TDEs, and only black holes with rapid spin can launch relativistic jets during TDEs. Further modeling suggests that the jet model can successfully account for the X-ray \citep{Lei13,Tch14} and the radio \citep{Metzger12,Wang14,Liu15} data.

Table \ref{tab:TDE} lists the data for all the TDE candidates. TDE flares last much longer than GRBs and SBOs. They usually have a relatively fast rising phase, reach and stay at the peak for some time, and then decay roughly with a power law $ L\propto t^{-5/3}$. The luminous state usually lasts for several months to one year \citep{Ree88}. In our sample, only two events may have been detected both in the rising and declining phases, so that the peak luminosity was measured \citep{KB99,Esq08}. For other TDEs, one did not detect the sources in both the rising and the declining phases, so that the peak luminosity cannot be well constrained. The maximum luminosity during the detected phase is only the lower limit of the peak luminosity. However, since the lightcurve peak is rounded and spreads in several weeks \citep{Lodato09}, the observed peak luminosity may not be too different from the true peak luminosity.

Due to the narrow bandpass of the X-ray telescopes that detect TDEs, the spectral shape for TDEs is not well constrained. Except for the BAT detected TDEs for which a treatment similar to GRBs can be applied, for the majority of TDEs, we only apply an empirical relation to estimate the bolometric luminosity by multiplying X-ray luminosity by a factor of 1.4, i.e. \citep{Mak10}
\begin{equation}
L_{p,\rm bol}=1.4\times L_{p}.
\label{eq:k2}
\end{equation}
For Swift TDEs, a power-law spectrum in the BAT band is reported for both events with a photon index of 1.8 and 1.6, respectively \citep{Bur11,Cen12}. Even though the spectrum of jetted TDEs is not known, we speculate that they have a non-thermal spectrum with $E_p$ not far above the BAT band. As a result, we apply Eq.(\ref{eq:Ep-Gamma}) to estimate $E_p$ and apply Eq.(\ref{eq:kdf}) to estimate the $k$-correction factor assuming the standard Band-function parameters for the spectrum. This led to a $k$-correction factor 2.1, and 2.2 for Sw J1644+57 and Sw J2058+05, respectively.

\section{Results}
\subsection{LGRBs}
Since LGRBs and SGRBs have different physical origins (massive star core collapse vs. compact star mergers), we derive their event rate density and luminosity function separately. For each type, we first derive the GLF by ignoring possible redshift evolution effect. Then we dedicate one subsection to discuss the possible evolution effect. Within LGRBs, the LL-LGRBs have been claimed to form a distinct component in the luminosity function \citep{Liang07,Vir09}, which may have a somewhat different physical origin \citep[e.g.][]{Liang07,Bro11,NS12}. In our analysis, we adopt two approaches. First, we derive the luminosity function of HL-LGRBs and LL-LGRBs separately  \citep[e.g.][]{Liang07,Vir09}. In the second approach, we derive the luminosity function of LGRBs jointly by fitting the data together with a two-component (i.e. TPL) luminosity function. This approach is justified in view that both LL- and HL-LGRBs are associated with Type Ic supernovae, and therefore may share a common physical origin.

\subsubsection{Global luminosity function}

\begin{figure}[!htb]
\centering
\includegraphics[width=3in]{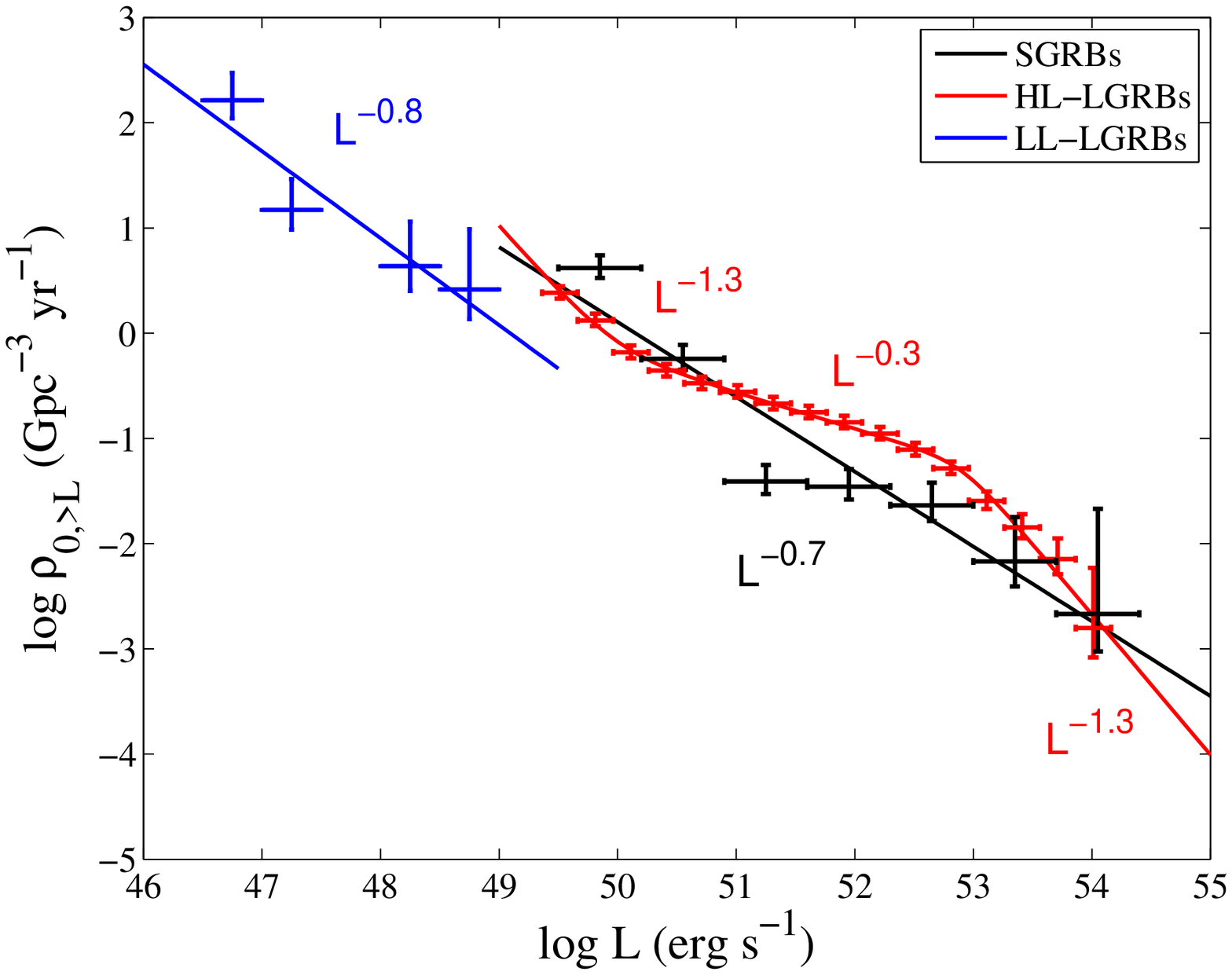}
\includegraphics[width=3in]{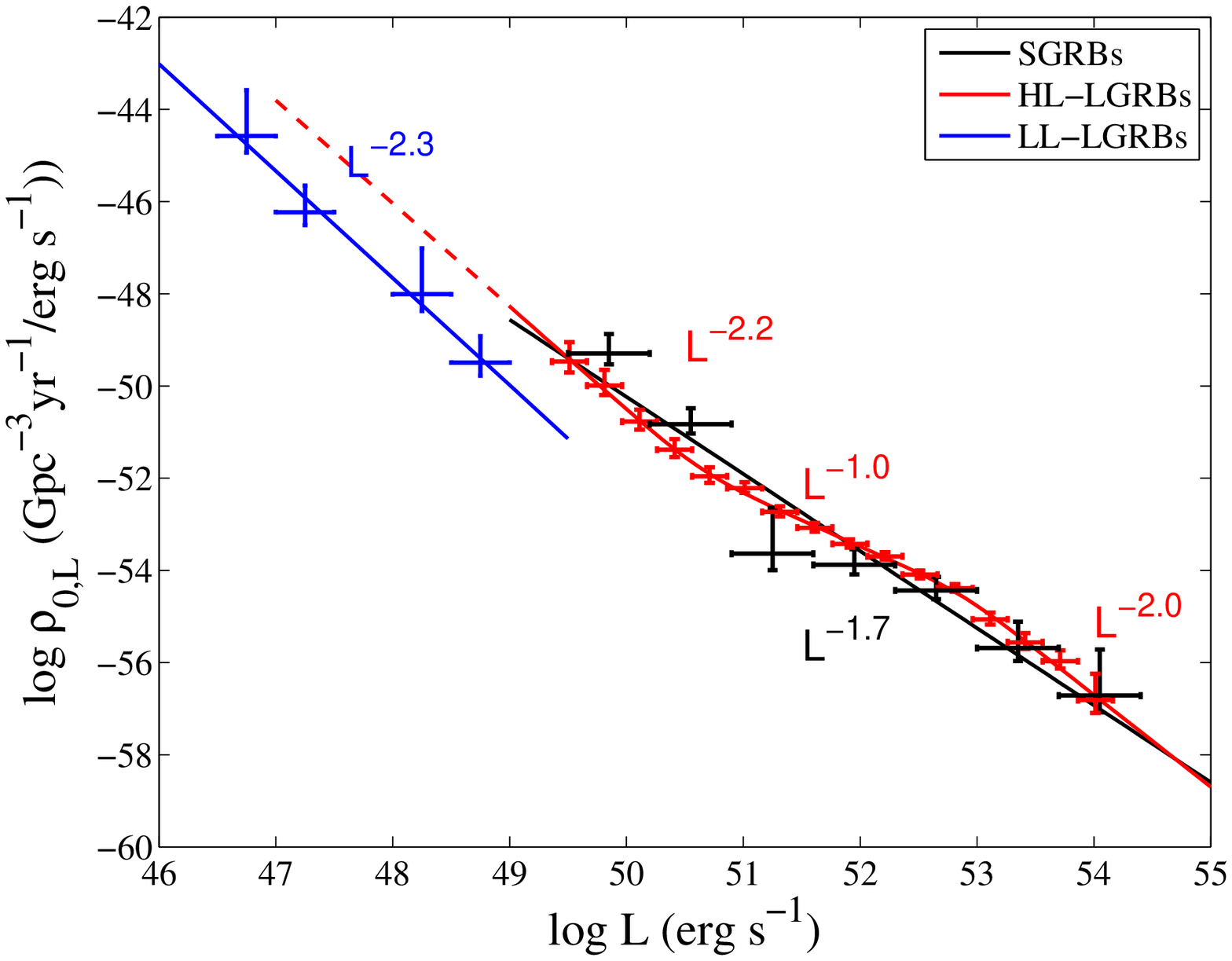}
\caption{Left panel: Event rate density ($ \rho_{0,>L} $) distribution for LL-LGRBs (blue), HL-LGRBs (red) and SGRBs (black). The luminosity bin has a width of 0.3 for HL-LGRBs, 0.5 for LL-LGRBs and 0.7 for SGRBs in the logrithmic space. For short GRBs, the Gaussian merger delay time model is adopted. The vertical error bars represent the $1 \sigma $ Gaussian errors calculated from \citep{Geh86}. The horizontal error bars show the width of the luminosity bin. Right panel: Luminosity functions of LL-LGRBs, HL-LGRBs, and SGRBs. LL-LGRBs and short GRBs can be fit with a single power law, with indices 2.3 and 1.7, respectively. HL-LGRBs are fit with a triple power law with $\alpha_1=2.2$ $ \alpha_2=1.0$ and $\alpha_3=2.0 $.} 
\label{fig:grb}
\end{figure}

Figure \ref{fig:grb} shows the results of GRBs. The event rate density above a particular $L$ as a function of the bolometric luminosity $L$ is presented in the left panel, and the specific event rate density as a function of $L$, which describes the GLF, is presented in the right panel. In both panels, the HL-LGRBs (red), LL-LGRBs (blue), and SGRBs (black) are presented separately. The luminosity bin is taken as 0.3 (LGRBs) or 0.7 (SGRBs) in the logarithmic space. The horizontal errors denote the luminosity bins, whereas the vertical errors are calculated from small-sample statistics \citep{Geh86}. The best fit lines for all three sub-types of GRBs are also presented. The fitting results are summarized in Table \ref{tab:LF} and Table \ref{tab:rho}. For all the fitting parameters, we also present the 1$\sigma$ range of errors based on 5000 sets of Monte Carlo simulations.

The local event rate density of LL-LGRBs is $ 164^{+98}_{-65} $ $\rm Gpc^{-3}\,yr^{-1}$ with a minimum luminosity $\rm 5\times10^{46}\,erg\,s^{-1}$, which is roughly consistent with previous results (\citealt{Liang07},\citealt{Vir09}). The longer working period of {\em Swift} till now than the ones relevant for the previous two papers makes the event rate density slightly lower than before.  However, it is still around two orders of magnitudes higher than that of HL-LGRBs, which is $\rm 2.4^{+0.3}_{-0.3} \,Gpc^{-3}\,yr^{-1}$ above $\rm 3\times10^{49}\,erg\,s^{-1}$, or $\rm 0.8^{+0.1}_{-0.1} \,Gpc^{-3}\,yr^{-1}$ above $\rm 10^{50}\,erg\,s^{-1}$ (a typical luminosity threshold adopted before). This local event rate of HL-LGRBs is slightly lower than \cite{Liang07} and \cite{Wan10}. 

\begin{figure}[!htb]
\centering
\includegraphics[width=4.5in]{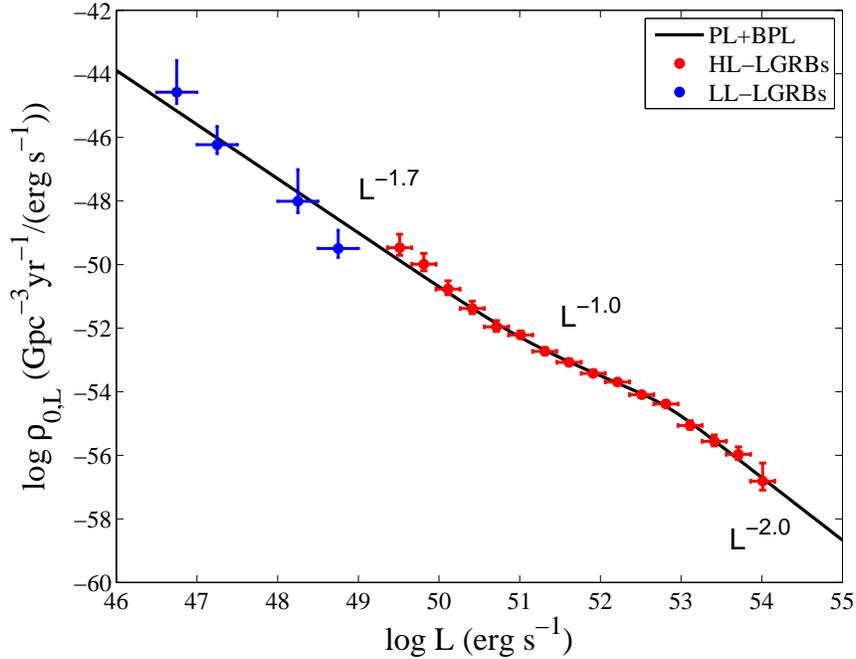}
\caption{A joint fit of LL- and HL-LGRBs with a two-component luminosity function.}
\label{fig:lgrb}
\end{figure}

From Eq.(\ref{eq:rho_0L}), one can see that $\rho_{0,L}$ is proportional to the luminosity function $\Phi(L)$, with $ \rho_0 $ defining the normalization. We use either a single power law (SPL) or a triple power law (TPL) to fit the $\rho_{\rm 0,L}$ distribution for each sub-type of GRBs to characterize their GLFs. The fitting results are shown in Table \ref{tab:LF}. HL-LGRBs can be fit with a TPL with $ \alpha^{\rm HL}_1=2.2^{+0.4}_{-0.2}$, $ \alpha^{\rm HL}_2=1.0^{+0.1}_{-0.1}$, $ \alpha^{\rm HL}_3=2.0^{+0.3}_{-0.3}$ and the break luminosity $ L_{b,1}^{\rm HL}=5.0^{+3.0}_{-1.3} \times 10^{50}$ $ \rm erg~s^{-1} $, $ L_{b,2}^{\rm HL}=7.1^{+4.3}_{-3.0} \times 10^{52}$ $ \rm erg~s^{-1} $. The luminosity distribution of LL-LGRBs can only be fit using a SPL with $ \alpha^{\rm LL}= 2.3^{+0.2}_{-0.2}$. The LL-LGRBs are not the straightforward extension of HL-LGRBs to low luminosities. But the slope of LL-LGRB is similar to that of first component of HL-LGRBs. The normalization of LL-LGRBs is a little bit lower as we apply a lower threshold for LL-LGRBs than HL-LGRBs. Therefore it may be possible that LL-LGRBs follow the extension of LF of HL-LGRBs if we take a TPL fit to the joint LL- and HL- samples. Such a fit is presented in Fig.\ref{fig:lgrb}, with the best fit parameters being $ \alpha_1=1.7^{+0.1}_{0.1}$, $ \alpha_2=1.0^{+0.2}_{-0.1}$, $ \alpha_3=2.0^{+0.3}_{-0.3}$ and the break luminosities $ L_{b,1}=1.0^{+0.2}_{-0.3}\times 10^{51}$ $ \rm erg~s^{-1} $, $ L_{b,2}=7.8^{+2.3}_{-3.1}\times 10^{52}$ $ \rm erg~s^{-1} $. This is similar to the TPL fit to HL-LGRBs alone except for a slightly shallower $\alpha_1$, which is compromised by the slight mis-match between LL-LGRBs and HL-LGRBs.

The TPL nature of the joint GLF for LGRBs is interesting. Such a feature was only noticeable in the past when LL-LGRBs are included \citep{Liang07,Vir09}. With the current sample, we find that it is required even for HL-LGRBs alone. With the joint fit LL- and HL-LGRBs, we find that the steep component in the low-luminosity end now extends all the way to $\sim 10^{51}~{\rm erg~s^{-1}}$, so that no clear separation between the LL- and HL-LGRB population is seen.

We would like to stress that if we focus on the two high-$L$ segments in the GLF, our results ($\alpha_2 \sim 1.0$, $\alpha_3 \sim 2.0$, $L_{b,2} \sim 7.8\times 10^{52}~{\rm erg~s^{-1}}$) are broadly consistent with previous results: $ \alpha^{\rm HL} \sim 0.8$, $\beta^{\rm HL} \sim 2.6 $, and  $ L_{b}^{\rm HL} \sim 2.5\times 10^{52}$ by \cite{Liang07}; and $ \alpha^{\rm HL} \sim 1.17$, $ \beta^{\rm HL} \sim 2.44 $, and  $ L_{b,2}^{\rm HL} \sim 3.1\times 10^{52}$ $ \rm erg\,s^{-s} $ by \cite{Wan10}\footnote{Notice $\alpha^{\rm HL}$ and $\beta^{\rm HL}$ in previous works correspond to $\alpha_2$ and $\alpha_3$ in our notation. Also \cite{Wan10} derived the luminosity function in the logarithmic space, so the $\alpha$ and $\beta$ values in their notation are smaller by one from our values.}.

According to Eqs.(\ref{eq:dfrho_0L}) and (\ref{eq:rho0'}), the indices of $\rho_{\rm 0,>L}$ and $ \rho_{\rm 0,L}dL $ should be roughly the same, i.e. the index of $\rho_{\rm 0,>L}$ should be greater than the index of $\rho_{\rm 0,L}$ by one. This is generally satisfied for most of the transients studied in this paper (see all the indices marked in Figs.3-8). 

\subsubsection{Luminosity function evolution}
\begin{figure}[!htb]
\centering
\includegraphics[width=3in]{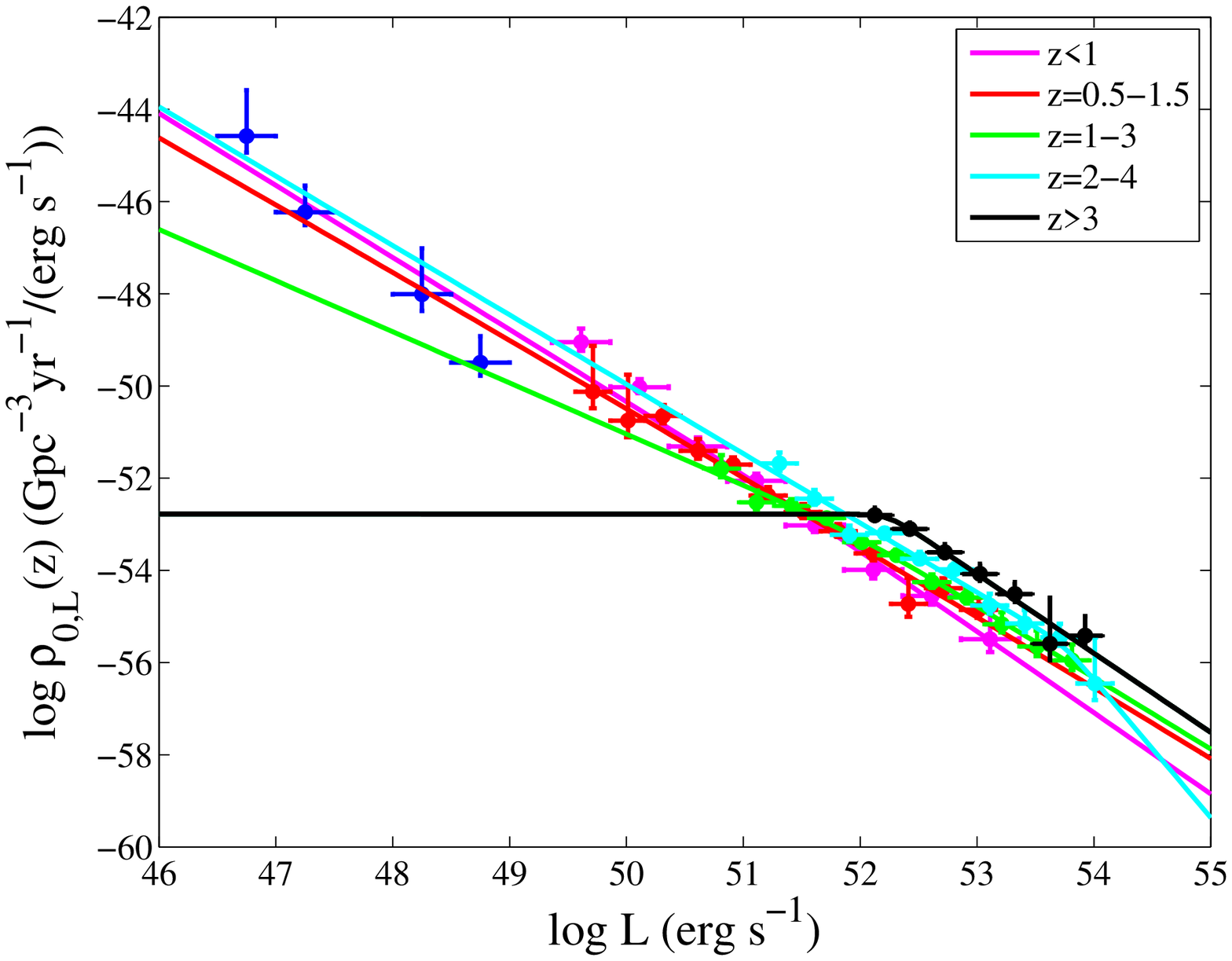}
\includegraphics[width=3in]{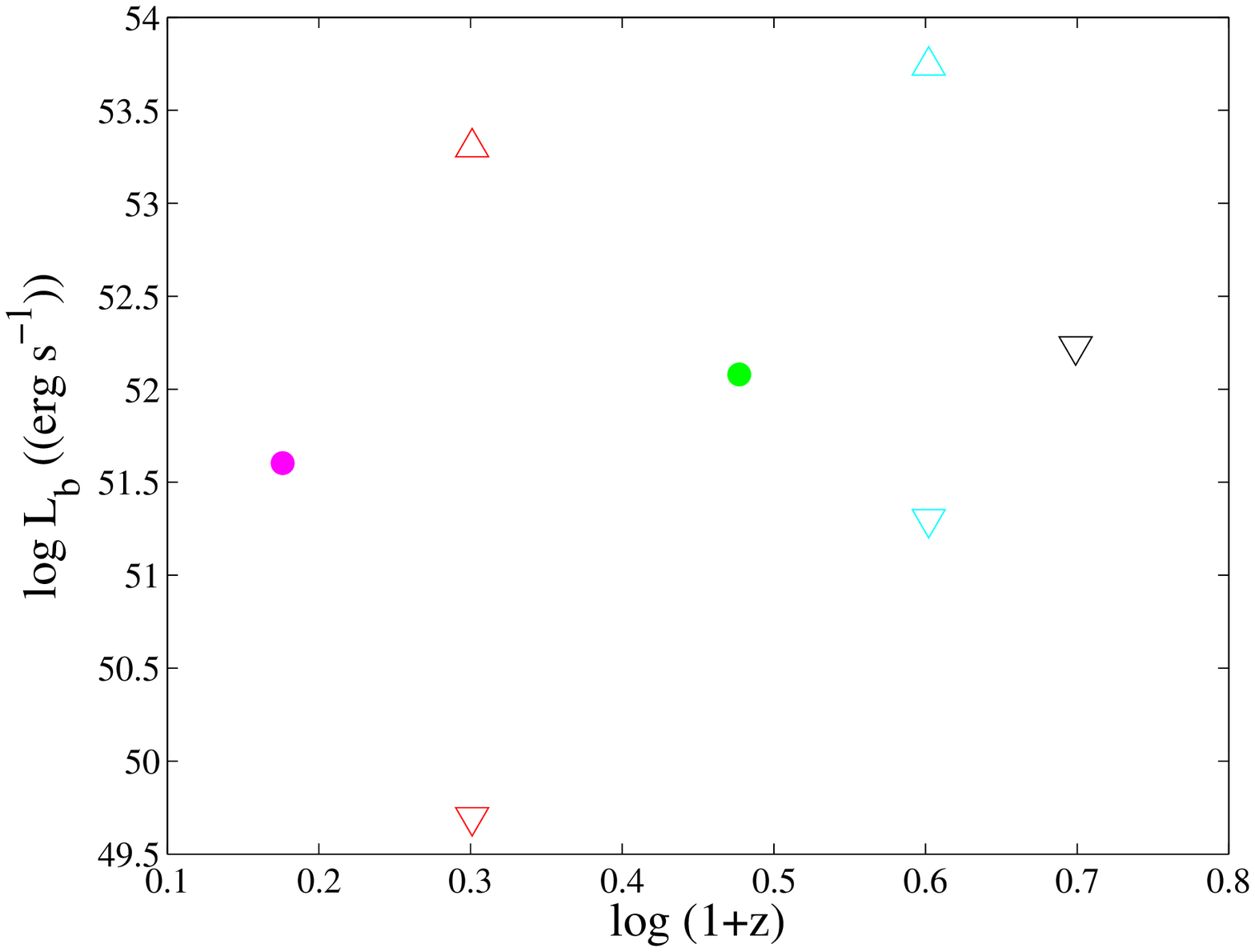}
\caption{Left panel: Luminosity function of long GRBs in different redshift bins for the full sample. Blue dots denote the LL-LGRBs. Magenta, red, green, cyan and black dots represent HL-LGRBs from $0<z<1$, $0.5<z<1.5$, $1<z<3$, $2<z<4$ and $z>3$, respectively. Best fit models are overplotted as solid curves with the corresponding color. Right panel: Break luminosity evolution inferred from the luminosity function fit from the left panel. The medium values of each redshift bin is taken. For dubious BPL, we give both upper limit (the minimum luminosity of data, lower triangle) and lower limit (derived from BPL fit, upper triangle). No clear evolution pattern is observed. } 
\label{fig:LF1}
\end{figure}

Using method laid out in Section \ref{sec:evolution}, we study the evolution effect of LGRB luminosity function. The results of $\rho^z_{\rm 0,L}$ are shown in Fig.\ref{fig:LF1} left panel. For each redshift bin, we fit the luminosity function with a SPL or BPL if the latter is needed. The fitting parameters are presented in Table \ref{tab:LFApp}. One can see that indeed there is an apparent luminosity function evolution effect. However, there is no clear pattern to quantify the evolution. The right panel of Fig.\ref{fig:LF1} shows the redshift-dependent break luminosity. In some redshift bins a break is clearly seen. However, in some other bins, the break either does not exist, or simply required by only one data point with low significance. For SPL fits, we place either a lower limit or an upper limit based on the highest or lowest luminosity bin. As shown in Fig.\ref{fig:LF1} right panel, there is no clear pattern to quantify the evolution effect.  Further more, the PL indices $\alpha_1$ and $\alpha_2$ also show significant evolutions (variations) in different redshift bins. Since different redshift bins have different $L_m$, and since the behavior below $L_m$ is poorly constrained by the data, in Table \ref{tab:LFApp} we choose different $L_m$ for different redshift bins. For nearby universe, we also get a $\rho^{z}_0$ at higher $L_m$. We obtain a consistency of $\rho^{z}_0$ derived from data from different redshift interval.

\begin{figure}[!htb]
\centering
\includegraphics[width=3in]{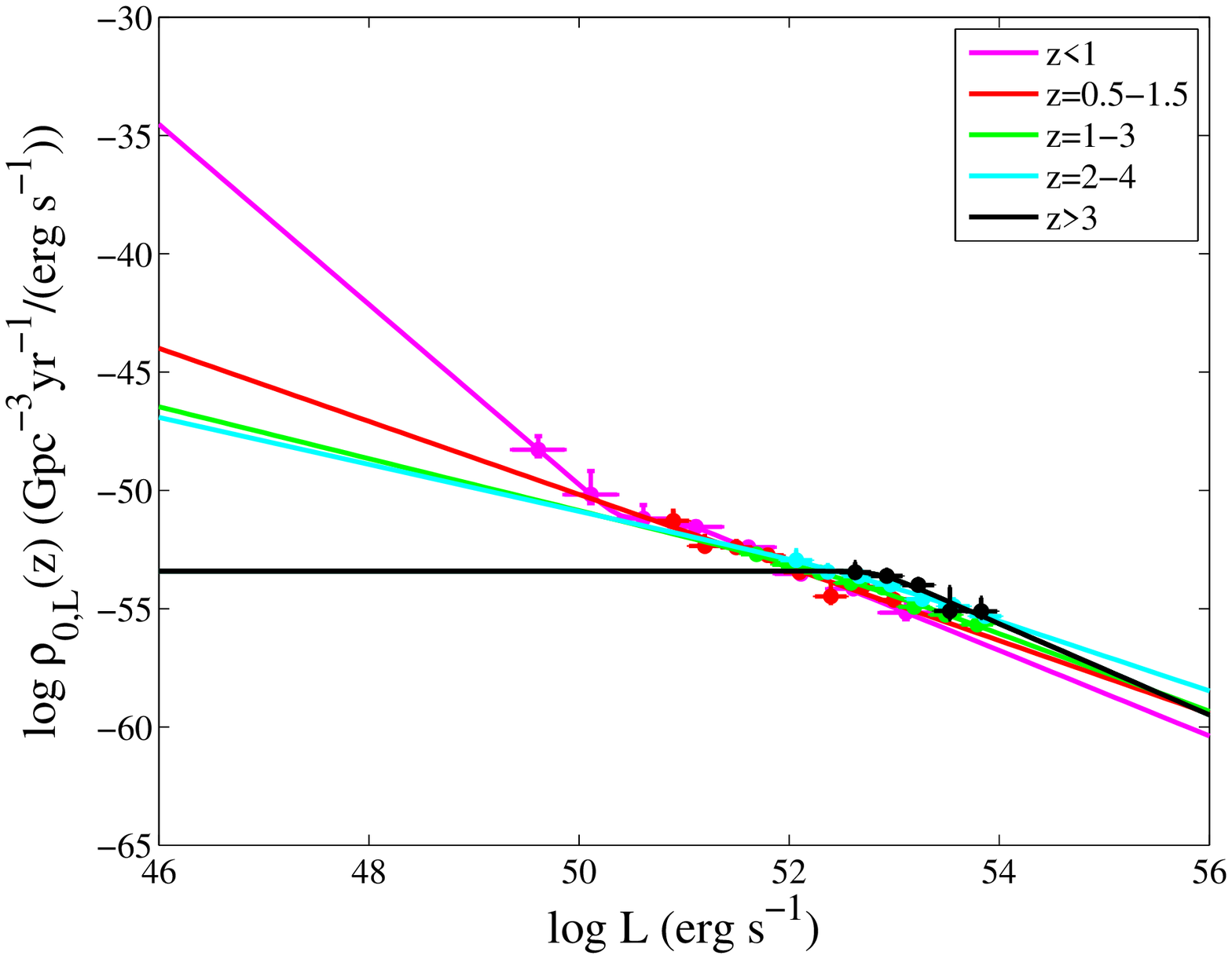}
\includegraphics[width=3in]{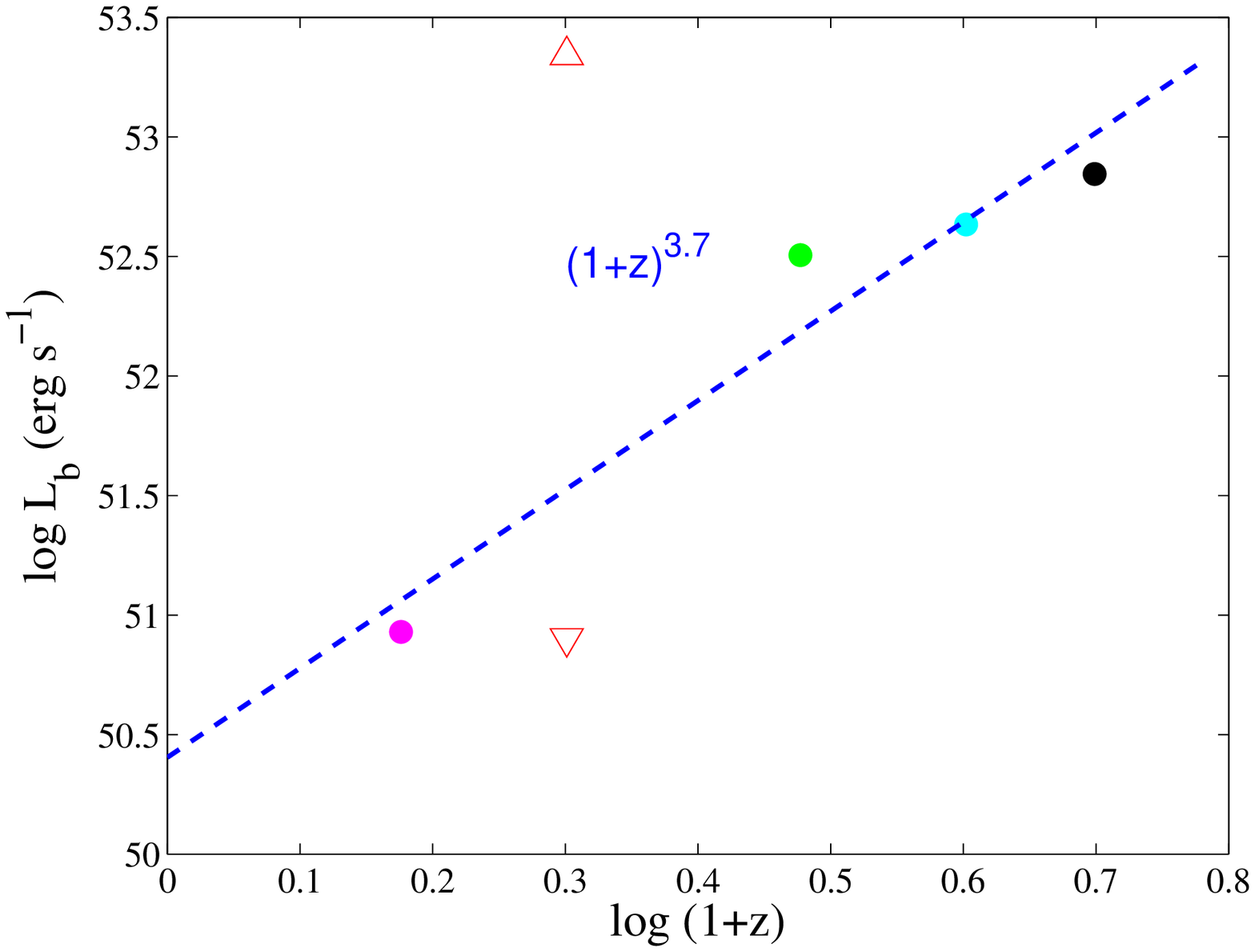}
\caption{Same as Fig.\ref{fig:LF1} but for a sub-sample with peak photon flux larger than $\rm 1.8~ph~s^{-1}~cm^{-2}$. In the redshift bin 0.5-1.5 (red), the sample is fit by a single power law. The lower/upper limits of the break luminosity (red triangles in the right panel) are derived from the maximum and minimum luminosities of the data. These limits are not used in the $ L_b $ fit (the dashed line in right panel). } 
\label{fig:LF2}
\end{figure}

To minimize the truncation effect by the flux limit of detectors, we also use a sub-sample with a higher threshold (with peak photon flux larger than 1.8$\rm~ph~s^{-1}~cm^{-2}$), see Fig.\ref{fig:LF2}. The sub-sample consists less GRBs in some redshift intervals (e.g. $ z\in (0.5,1.5) $) so that the luminosity function can be fit by a SPL. In the nearby universe, on the other hand, since LL-LGRBs are dropped due to this high-threshold criterion, the luminosity function demands a TPL fit\footnote{For the full sample, the inclusion of LL-LGRBs compensates the low-$L$ excess so that a BPL presents a reasonable fit.}.
The fitting parameters are also presented in Table \ref{tab:LFApp}. The evolution of $L_b$ is now better quantified by $L_b \propto (1+z)^{3.7}$ (Fig.\ref{fig:LF2} right panel), but a signficant variation of $\alpha_1$ and $\alpha_2$ values in different redshift bins remain. We therefore conclude that there is no straightforward way to quantify the the evolution effect of LGRB luminosity function.

\begin{figure}[!htb]
\centering
\includegraphics[width=3in]{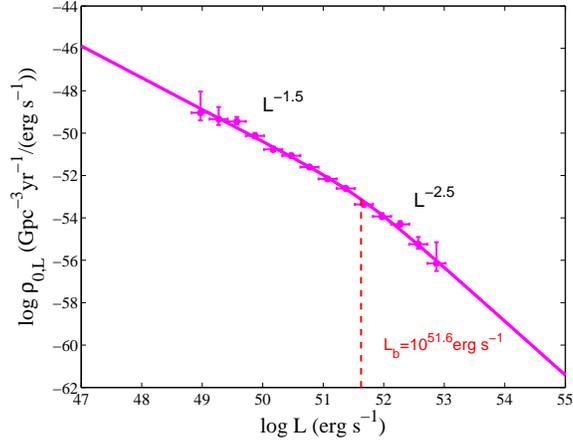}

\caption{The derived local luminosity function of HL-LGRBs assuming a simple luminosity function evolution model with same luminosity function shape but an evolving break luminosity $L_b \propto (1+z)^{2.3}$. The broken power law LF gives $\alpha_1=1.5$, $\alpha_2=2.5$ and $L_b=51.6$ $\rm erg~s^{-1}$.}
\label{fig:Levl}
\end{figure}

On the other hand, if we assume the evolution law assumed from the latest papers \citep[e.g.][]{Petro15,Yu15,Pescalli15}, i.e. the PL indices remain unchanged, and only $L_b$ evolves with $k\sim2.3$, we can map the luminosity function from the data by correcting the luminosity $L$ at $z$ to the ``local'' value, i.e. $L_0=L/(1+z)^{2.3}$, to derive the ``local'' luminosity function. The result is shown in Fig.\ref{fig:Levl}. This local luminosity function can be fit by a broken power law with $\alpha_1=1.5$, $\alpha_2=2.5$ and $L_{b,0}=51.6~\rm~erg~s^{-1}$. This is consistent with the results presented in previous papers.

\subsection{SGRBs}
\subsubsection{Global luminosity function}
The local event rate density for SGRBs vary slightly for different merger delay models. For a minimum luminosity $\rm 7\times10^{49}\,erg\,s^{-1}$, it is $ 4.2^{+1.3}_{-1.0} $, $ 3.9^{+1.2}_{-0.9} $, $ 7.1^{+2.2}_{-1.7} $ $\rm Gpc^{-3}\,yr^{-1}$ for the Gaussian, log-normal and power-law delay models, respectively. Taking a typical minimum luminosity $ 10^{50}~ {\rm erg~s^{-1}}$ as adopted by previous authors, the numbers are  $1.3^{+0.4}_{-0.3}$, $1.2^{+0.4}_{-0.3}$, and $3.3^{+1.0}_{-0.8}$ $\rm Gpc^{-3}\,yr^{-1}$, respectively. The local event rate density for the log-normal model is slightly lower than the value reported by \cite{Wan14}, as we use a slightly lower flux threshold for {\em Swift} BAT-detected SGRBs.

\begin{figure}[!htb]
\centering
\includegraphics[width=5in]{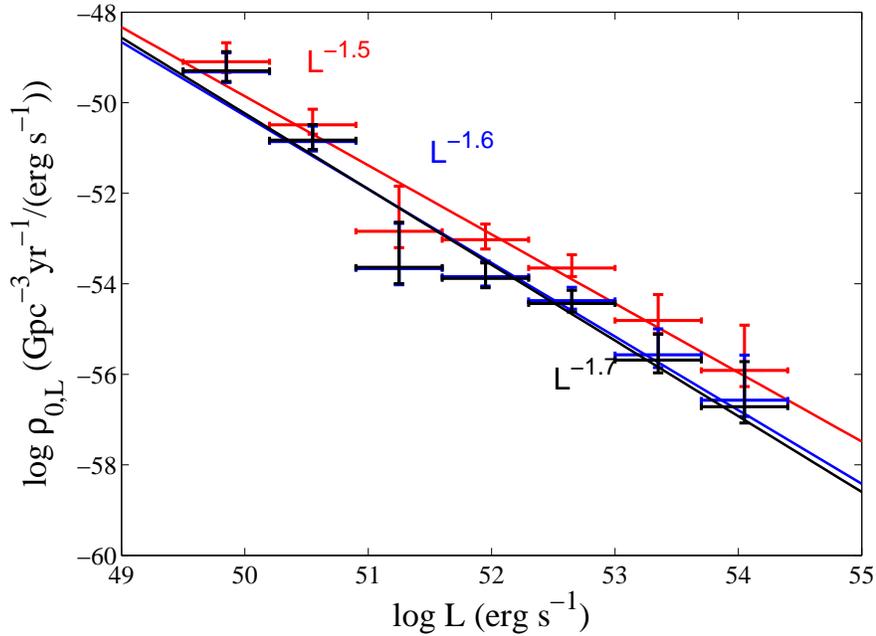}
\caption{Luminosity functions for SGRBs with three different merger delay models: Gaussian (black), log-normal (blue), and power-law (red).}
\label{fig:sgrb}
\end{figure}

SGRBs come with a much smaller number of events than LGRBs. Assuming that all the SGRBs with redshift measurements are of a compact star merger origin, we derive their luminosity function in Fig.\ref{fig:sgrb}. The three different merger delay models give slightly different results, but in general all three models are consistent with having a SPL luminosity function with an index of $\sim 1.6$ (for details, see Fig.\ref{tab:LF}). This is different from HL-LGRBs which require a TPL luminosity function. It is also different from \cite{Wan14}, who claimed a BPL luminosity function. 

One caveat of our treatment is the assumption that all SGRBs are of a compact star merger (Type I) origin. In the {\em Swift} era, \cite{Zhang09} first suspected that some high-$L$ SGRBs at high redshifts may not be of the Type I origin, but may rather originate from massive star core collapse (Type II). They suggested to apply multi-wavelength criteria (instead of using $\gamma$-ray duration only) to judge the physical origin of a GRB. Later \cite{Vir11} pointed out that the assumption that all SGRBs are of the compact star merger origin (Type I) is disfavored since one cannot simultaneously account for the $z-L$ two-dimensional distribution and the $\log N-\log P$ distribution of SGRBs. They claimed that a good fraction of SGRBs may be of a Type II origin. \cite{bromberg13} recently reached the similar conclusion using a different argument based on the duration distribution of GRBs. \cite{Wan14} used the criteria of \cite{bromberg13} and excluded about 1/3 of SGRBs that they suspect to be of a massive star core collase origin. This may explain the difference between the results of this paper (SPL luminosity function) and that paper (BPL). We did not exclude any SGRB from our sample for the following reason. We believe that any conclusion about the physical category of a SGRB based on the duration information only \citep[e.g.][]{bromberg13} is not reliable. Rather one should consider multi-wavelength criteria \citep{Zhang09}, especially the host galaxy type and the afterglow location within the host. Host galaxy studies of SGRBs \citep{fong10,berger14} suggested that the hypothesis that all SGRBs belong to the compact star merger (Type I) category is not in conflict with the data. Indeed, some SGRBs excluded by \cite{Wan14} (defined by them as collapsars) actually have large offsets from host galaxies, fully consistent with being due to a compact star merger origin (e.g. GRB 070724 with offset $5.46 \pm 0.14$ kpc, and GRB 070809 with offsef $33.22\pm 2.71$ kpc, \citealt{fong10,fong13}). Furthermore, a recent study of SGRB emission amplitude parameter \citep{Lv14} also suggested that most observed SGRBs are not disguised SGRBs due to the ``tip-of-iceberg'' effect. Even though we believe that there exists a contamination Type II GRBs in the SGRB sample, without studying the multi-wavelength data of SGRBs in detail, we believe that it is more appropriate not to exclude any SGRB in this study.

\subsubsection{Luminosity function evolution}
\begin{figure}[!htb]
\centering
\includegraphics[width=3in]{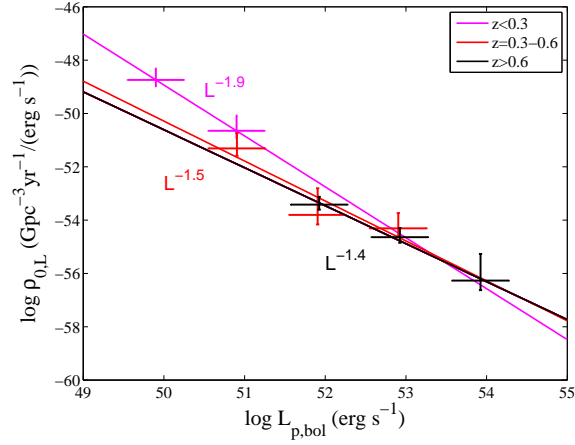}
\caption{Luminosity functions of short GRBs in three different redshift bins.} 
\label{fig:LFsS}
\end{figure}

For short GRBs that have much less data, it is more difficult to study the possible evolution effect of luminosity function. We manage to divide the short GRB sample into three redshift bins ($z<0.3$, $0.3<z<0.6$, and $z>0.6$), and apply the method in Section \ref{sec:evolution} to derive $\rm \rho^z_{0,>L_m}$ using the data in different redshift bins. The results are shown in Fig.\ref{fig:LFsS}. The luminosity functions can be all fit with a SPL. The slope in the first redshift bin $(0<z<0.3)$, $1.9^{+0.3}_{-0.3}$, is somewhat steeper than the those of other two redshift bins ($1.5^{+0.2}_{-0.2}$ for $0.3<z<0.6$; and $1.4^{+0.2}_{-0.2}$ for $z>0.6$, respectively). However, the slopes are consistent with each other within error.

\subsection{Shock breakouts}
\begin{figure}[!htb]
\centering
\includegraphics[width=3in]{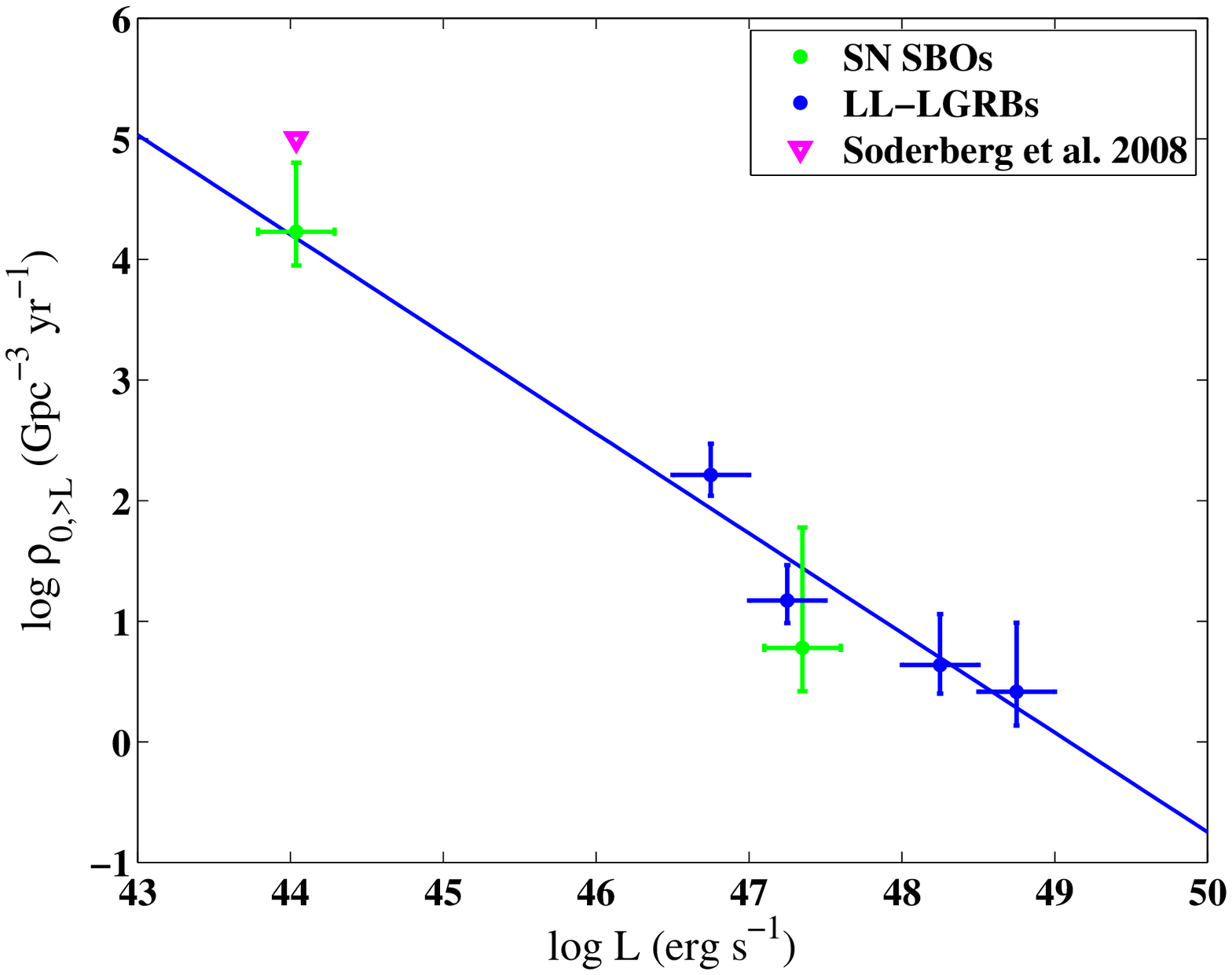}
\includegraphics[width=3in]{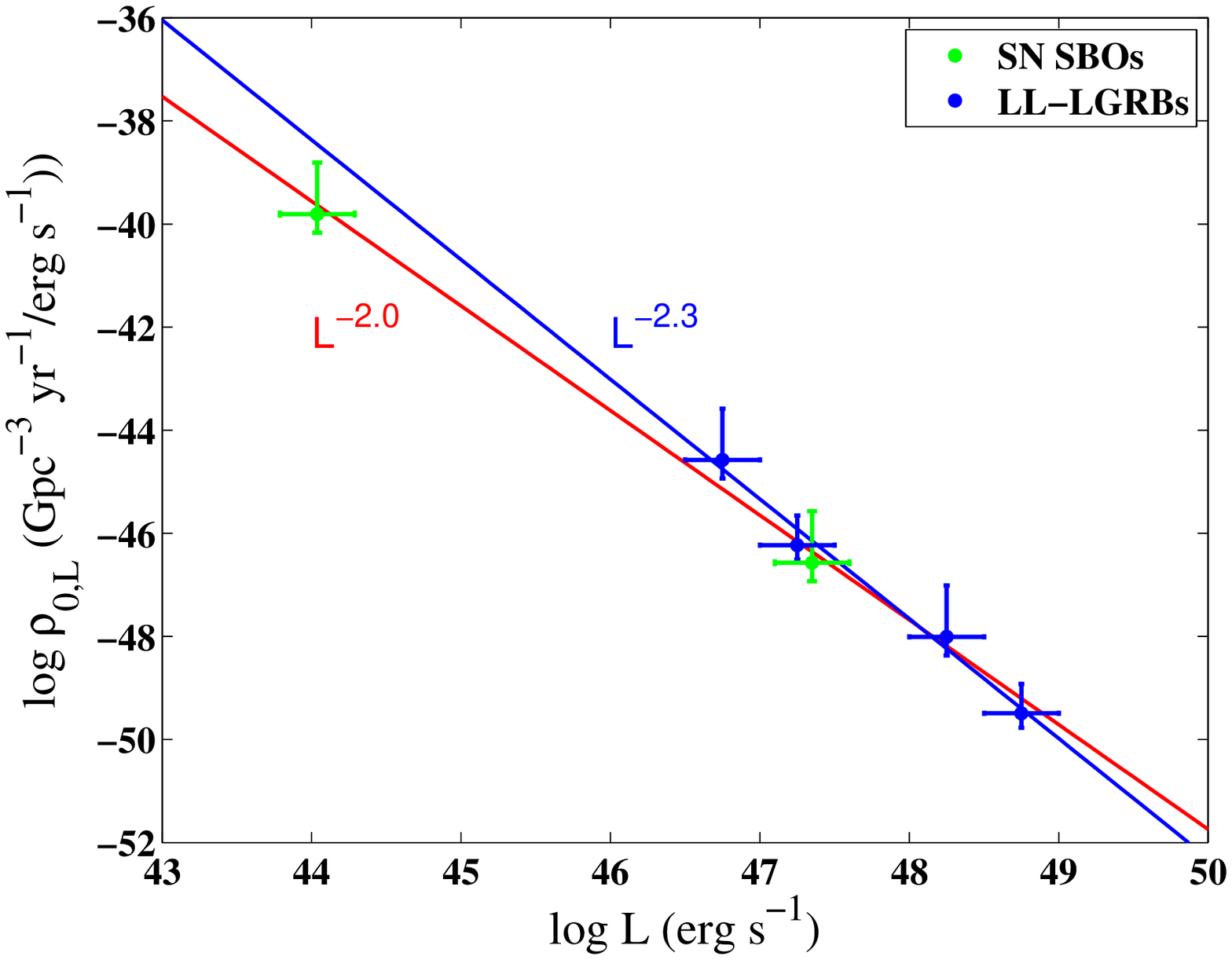}
\caption{Left panel: Event rate density ($ \rho_{0,>L} $) distribution for SBOs (green) and LL-LGRBs (blue). Right panel: Joint luminosity function of SBOs (green) and LL-GRBs (blue). For both panels, single power law fits to LL-LGRBs alone (blue) are shown. One can see that the SBO event XRO 080109/SN 2008D roughly follows the extension of the blue line. For the right panel, we also show the SBO/LL-LGRB joint-fit luminosity function (red). One can see that the slopes of the blue and red lines are similar to each other.}
\label{fig:sbo}
\end{figure}

The results for these two SBO events are presented in Fig.\ref{fig:sbo}. In view of the possible connection between SBOs and LL-LGRBs in general, we  present LL-LGRBs in the same plot for comparison. 
 
Our results suggest that the event rate density is $ 3.1^{+4.1}_{-2.0} \times 10^{4} $ $\rm Gpc^{-3}yr^{-1}$ for XRO 080109/SN 2008D-like SBO events (luminosity $\sim 6.1 \times 10^{43}~{\rm erg~s^{-1}}$), and is $11^{+25}_{-9} $ $\rm Gpc^{-3}yr^{-1}$ for GRB 060218/SN 2006aj-like SBO events (luminosity $\sim 1.5 \times 10^{47}~{\rm erg~s^{-1}}$). The former implies that the local event rate density of SBO is at least $ 10^4 $ times higher than that HL-LGRBs. A sensitive large field-of-view X-ray detector would lead to discovery of a large sample of these events. 

For both figures we find that XRO 080109/SN 2008D (green data point at low luminosity) follows the extension of LL-LGRBs (blue line). We also perform a joint fit between SBOs and LL-LGRBs and get a SPL luminosity function with slope $\sim 2.0$, which is similar to that of the slope of LL-LGRBs only ($ \alpha^{\rm LL}= 2.3$). This lends support to the possible connection between LL-LGRBs and SBOs \citep[e.g.][]{wangxy07,NS12,BD14}.

\subsection{Tidal disruption events}
\begin{figure}[!htb]
\centering
\includegraphics[width=3in]{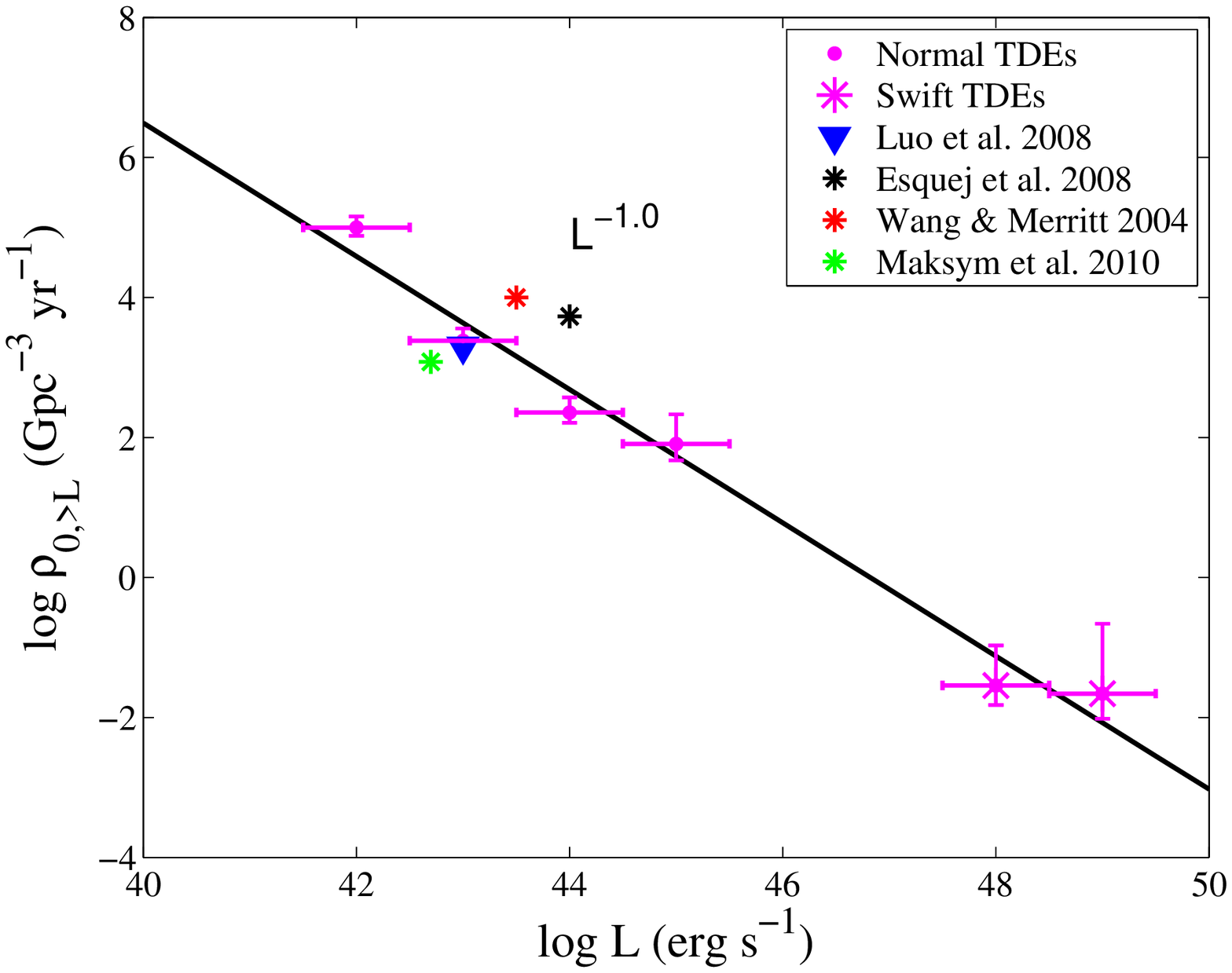}
\includegraphics[width=3in]{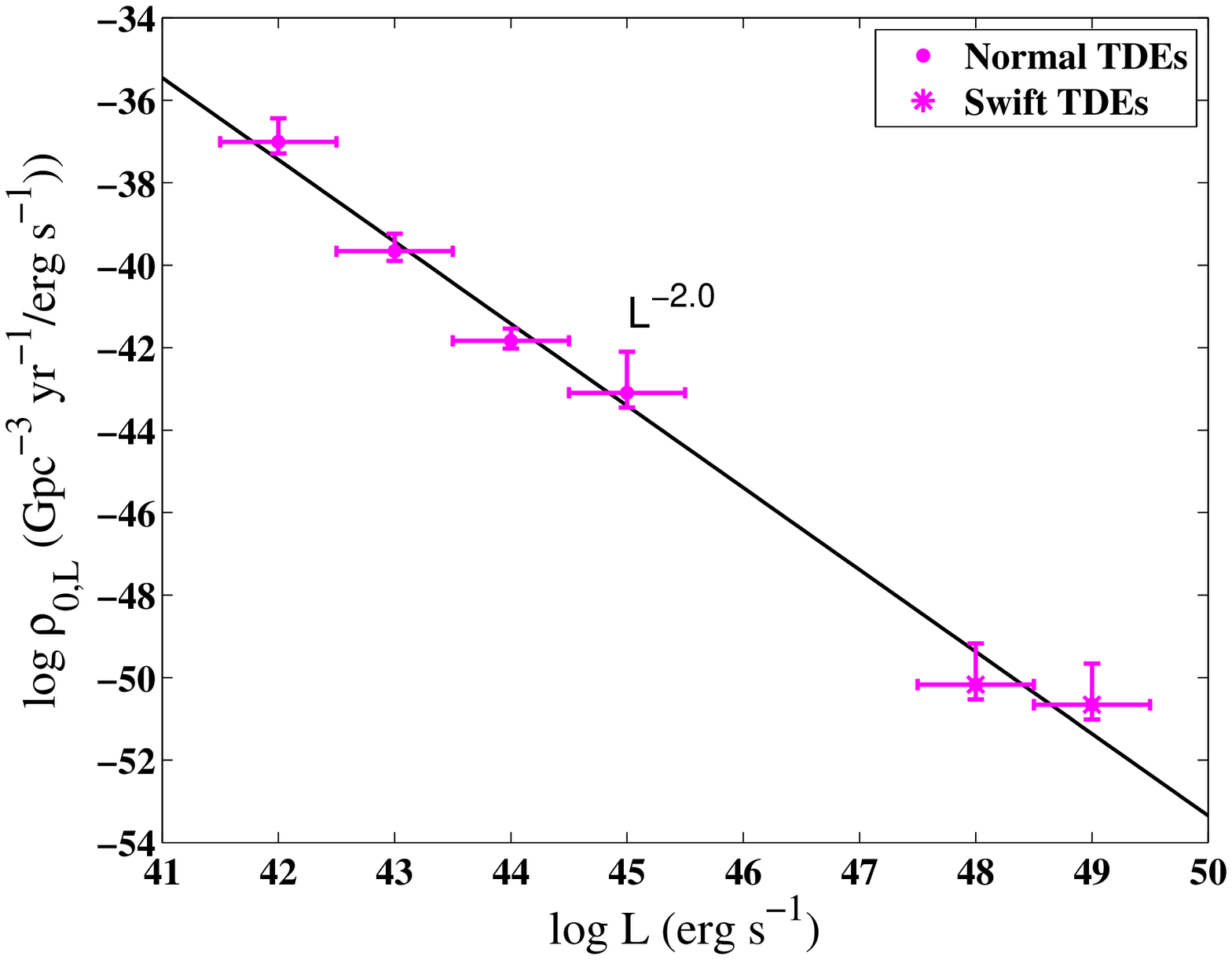}
\caption{Left panel: Event rate density ($ \rho_{0,>L} $) distribution for TDEs. The luminosity bin has a width of 1.0.
Several results from previous works are also shown for comparison. Right panel: Luminosity function of TDEs with the best fit. }
\label{fig_tde}
\end{figure}

Figure \ref{fig_tde} presents the results for TDEs. As mentioned above, two caveats are that the measured luminosities are only the lower limits of the ``peak luminosities'', and that some uncertainties are associated with the $k$-correction parameters. Bearing in mind these caveats, the following conclusions can be drawn. The event rate densities at different luminosity bins have a wide distribution, ranging from over $\rm 10^5~Gpc^{-3}~yr^{-1} $ at $ \rm 10^{42}~erg~s^{-1} $ to $\sim \rm 10^2~Gpc^{-3}~yr^{-1} $ at $ \rm 10^{45}~erg~s^{-1} $. In the luminosity range of $ (10^{43}-10^{44})~ \rm erg\,s^{-1}$, the event rate density is in the range of $ (10^3-10^4) $ $\rm Gpc^{-3}~yr^{-1} $, which is consistent with both theoretical predictions \citep{WM04} and the estimates based on observations \citep{Esq08,Luo08,Mak10}. The event rate density of the two {\em Swift} TDEs is $\sim \rm 0.03^{+0.04}_{-0.02} ~Gpc^{-3}yr^{-1} $. Similar to GRB 060218, these two events were detected through image triggers. In particular, Sw J1644+57 triggered BAT multiple times (all image triggers) \citep{Bur11}. We use the brightest peak to define the peak luminosity, and derive the event rate density based on the trigger information of that epoch.

The GLF of TDEs (including both normal TDEs and jetted TDEs) can be roughly described by a single PL with $ \alpha^{\rm TDE}=2.0 $. With the current sample, the jetted TDEs detected by {\em Swift} seem to lie in the extension of normal TDE luminosity function to high-luminosities. The event rate density of Sw J1644+57 shows a flattening feature at the highest luminosity. More data are needed to verify whether jetted TDEs form a distinct component in the global TDE luminosity function. 

\subsection{Global distribution}

\begin{figure}
\includegraphics[width=3.2in]{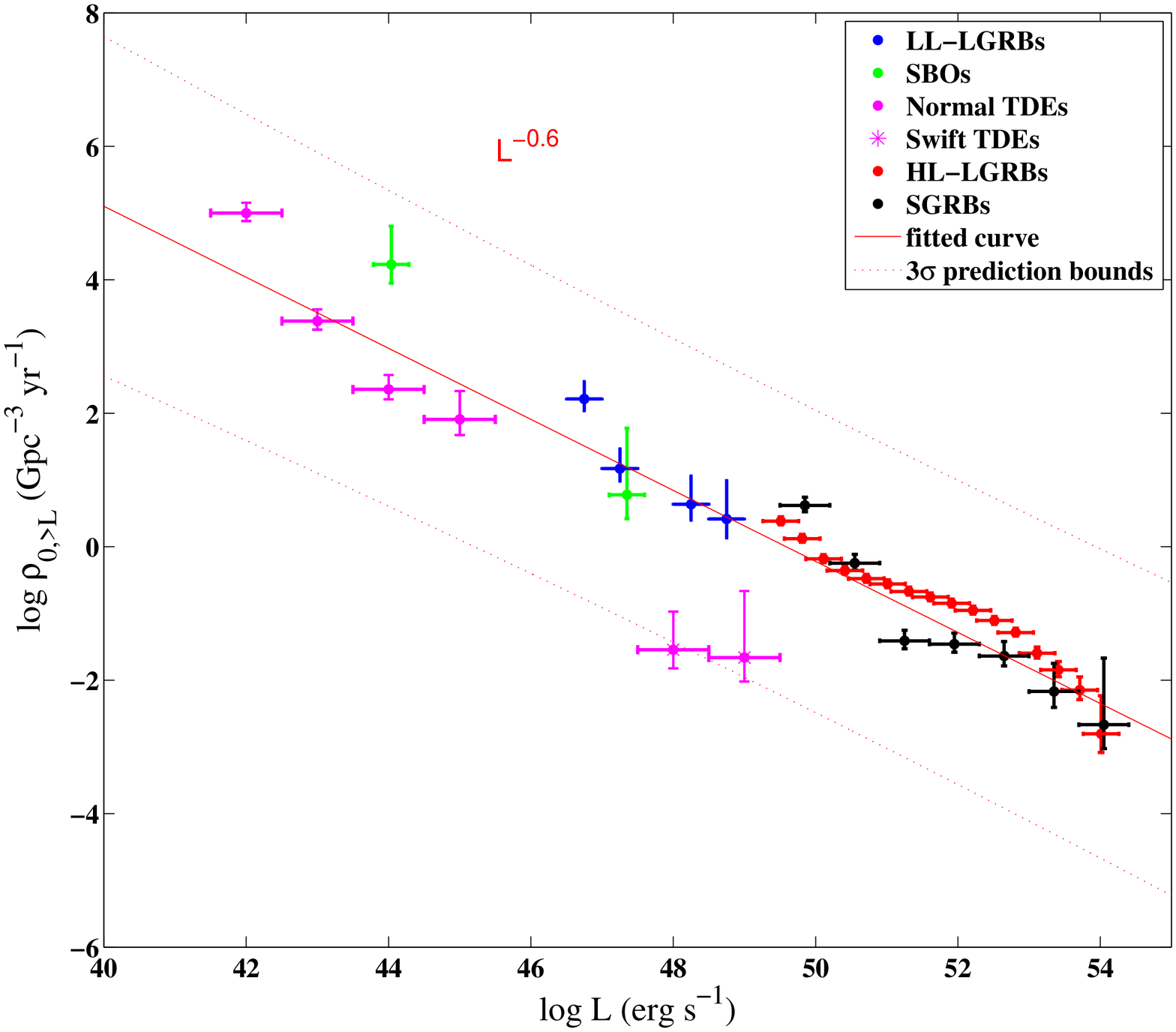}
\includegraphics[width=3.2in]{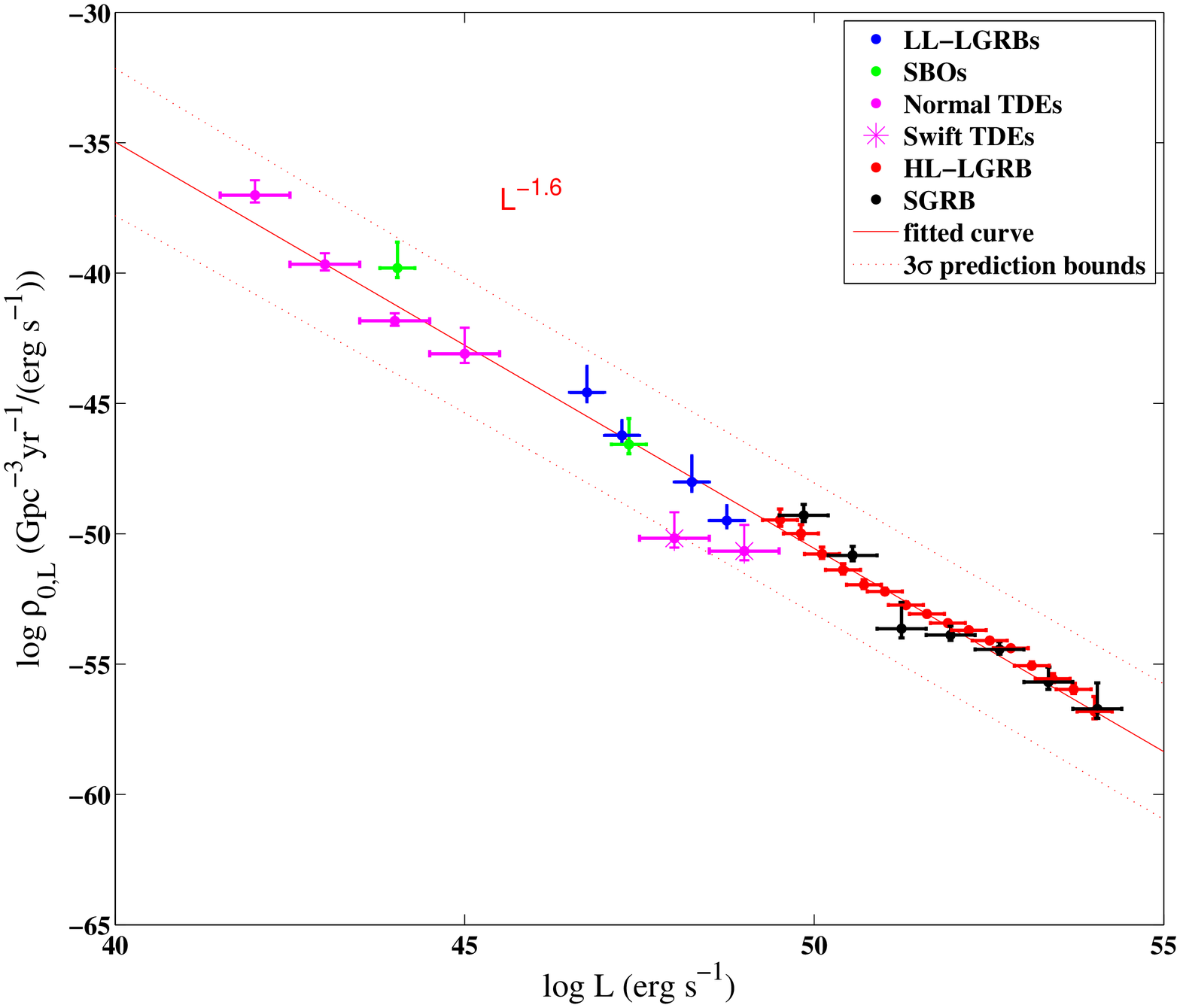}
\caption{Glocal distributions of all extra-galactic high-energy transients. Left panel: Event rate density above a minimum luminosity with respect to bolometric luminosity; Right panel: Joint luminosity function. Different types of events are marked in different colors. A single power law fit (red line) and $3\sigma$ boundary for the correlation are presented.
\label{fig:glb}}
\end{figure}

In Fig.\ref{fig:glb}, we plot all high-energy transients in one figure. The left panel shows the local event rate density above a certain luminosity ($\rho_{0,>L}$) as a function of $L$, and the right pabel presents the specific local event rate density ($\rho_{\rm 0,L}$), which essentially represents the GLF. All the data points are given a $1\sigma$ error in vertical axis. The width of luminosity bin is shown as the horizontal error bar.

Intriguingly, all the transients seem to line up to form a rough single power law distribution. A best fit to the GLFs of all the transients gives a slope of $\alpha^{\rm global}=1.6 $. All the events lie within the $3\sigma$ confidence bounds of the best fit. The region below this correlation line is likely due to an observational bias, and could be filled with new types of transients. With a much lower event rate density, these transients may not have been detected within the time span of the modern high-energy astronomy. The region above the correlation line, on the other hand, is not subject to selection effects and must be intrinsic.  The existence of such an upper boundary of high-energy transients is intriguing, which may be rooted from more profound physical reasons.

\section{Conclusions and discussion}

In this paper, we systematically investigated the local event rate density, redshift evolution, and GLF of several known extra-galactic high-energy ($\gamma$-ray and X-ray) transients, including high-luminosity and low-luminosity long GRBs that have a massive-star core-collapse origin, short GRBs that likely have a compact star merger origin, supernova shock breakouts, and tidal disruption events of stars by super-massive black holes. Our conclusions can be summarized as follows.
\begin{itemize}
 \item For all types of transients, the GLFs are typically well described by a single power law, although that of HL-LGRBs demands a triple power law.
 \item The local event rate density of each type of transient depends on the minimum luminosity. For GRBs, we get $\rho_{0,>L_m}^{\rm LL} = 164^{+98}_{-65}$ $\rm Gpc^{-3}yr^{-1}$ for $ L_m^{\rm LL}= 5\times 10^{46}$ $\rm erg\,s^{-1}$ for LL-LGRBs; $ \rho_{0,>L_m}^{\rm HL}=2.4^{+0.3}_{-0.3} $ $\rm Gpc^{-3}yr^{-1}$ with $ L^{\rm HL}_m= 3 \times 10^{49}$ $\rm erg\,s^{-1}$) for HL-LGRBs, and $ \rho_{0,>L_m}^{\rm SGRB}=4.2^{+1.3}_{-1.0},~3.9^{+1.2}_{-0.9},~7.1^{+2.2}_{-1.7}$ $\rm Gpc^{-3}yr^{-1}$ for SGRBs, with $ L^{\rm SGRB}_m=7 \times 10^{49} $ $\rm erg\,s^{-1}$ for the Gaussian, log-normal and power law merger delay models, respectively. Even though with two confirmed cases, the SBOs have event rate densities cover a wide range, from $ 3.1^{+4.1}_{-2.0} \times 10^{4} $ $\rm Gpc^{-3}yr^{-1}$ for XRO 080109/SN 2008D-like events (luminosty $\sim 6.1 \times 10^{43}~{\rm erg~s^{-1}}$) to  $ 11^{+25}_{-9} $ $\rm Gpc^{-3}yr^{-1}$ for GRB 060218/SN 2006aj-like events (luminosity $\sim 1.5 \times 10^{47}~{\rm erg~s^{-1}}$). The event rate density of TDEs also covers a wide range, from $ 1.0^{+0.4}_{-0.3}\times 10^{5}$ $ \rm Gpc^{-3}yr^{-1}$ with $ L_m^{\rm TDE}=10^{42} $ $\rm erg\,s^{-1}$ for normal TDEs to  $0.03^{+0.04}_{-0.02} $ $ \rm Gpc^{-3}yr^{-1}$ above $10^{48} $ $\rm erg\,s^{-1}$ for jetted TDEs detected by {\em Swift}.
 \item For GRBs, we confirmed the previous work \citep{Liang07,Vir09} that LL-LGRBs still do not straightforwardly follow the extension from HL-LGRBs. However, a TPL fit to the entire LGRB population suggests that the steeper GLF slope in the low-energy end now extends to a much higher luminosity, so that LL- and HL-LGRBs are no longer clearly separated. Consider the GLF shape in the high-$L$ end, the indices ($ \alpha=1.0,~\beta=2.0 $) and the break luminosity ($L_{b,2}=7.8 \times 10^{52}~{\rm erg~s^{-1}}$) are generally consistent with (even though not identical to) what was found in previous work \citep{Liang07,Wan10}. For SGRBs, we found a SPL luminosity function with $\alpha^{\rm SGRB}=1.6$, in contrast with the BPL distribution found by \cite{Wan14}. The discrepancy may lie in different sample selection criteria: whereas \cite{Wan14} excluded about 1/3 of SGRBs, we included all the SGRBs in our analysis. 
 \item We confirm the conclusion of previous authors \citep[e.g.][]{Petro15,Yu15,Pescalli15} that the luminosity function of LGRBs likely evolve with redshift. However, we find that the evolution cannot be easily quantified with a simple analytical model. Nonetheless, if one assumes that shape of the LF does not change and only the break luminosity evolves with redshift, the consistent results as previous authors can be achieved.
 \item SBOs and LL-LGRBs have a similar index $\alpha^{\rm SBO}=\alpha^{\rm SBO/LL}=2.0$, supporting the idea that LL-LGRBs may be related to shock breakouts \citep{Cam06,Wax07,wangxy07,Bro11,NS12}.
 \item The global luminosity function of TDEs is consistent with a single power law with $\alpha^{\rm TDE}=2.0$. The jetted TDEs discovered by Swift seem to be consistent with the extension of normal TDEs to high-luminosty regime, even though a flattening feature is seen. More data are needed to judge whether jetted TDEs form a new component in the GLF.
 \item Intriguingly, all the high-energy transients are consistent with having a global single power-law distribution of GLFs with a slope 1.6. Even though there could exist transients below the line which have not been discovered, the lack of events above the line is real. The existence of such an upper boundary is intriguing, and its physical origin is unknown.
 \item To perform this analysis, we adopted/derived the redshift distribution factor $f(z)$ of various types of transients based on different models. For LGRBs and SBOs, we assume that the event rate density traces the SFH, and adopt the empirical model (Eq.(\ref{eq:sfh})) of \cite{Yuk08}. For SGRBs, through Monte Carlo simulations, we derived empirical $f(z)$ functions for three merger delay models: Eqs.(\ref{eq:gaussian}), (\ref{eq:ln}), and (\ref{eq:pl}) for the Gaussian, log-normal, and power-law, respectively. For TDEs, we assume that the event rate is constant within each galaxy, and derived an empirical formula of $f(z)$ (Eq.(\ref{eq:tde})) based on the black hole number density evolution following \cite{Sha13}. These empirical formulae can be directly used in the future.
\end{itemize}

This paper focuses on high-energy transients only, serving as a reference for the wide-field $\gamma$-ray astronomy and the upcoming wide-field X-ray astronomy (led by e.g. Einstein Probe, \citealt{Yuan15}). We notice that the phenomenology of all the transients studied in this paper extends to lower frequencies in the electromagnetic spectrum. For example, GRBs have multi-wavelength afterglows. SBOs can peak in the UV or even optical band if the progenitor star is large enough. UV and optical TDEs and radio counterparts of jetted TDEs have been discovered. The study of high-energy transients in the low-frequency domain is beyond the scope of this paper.

\acknowledgments

We thank an anonymous referee for constructive comments and suggestions, Takanori Sakamoto for important discussions on {\em Swift} trigger algorithms, Francisco J. Virgili for important communications on Monte Carlo simulations,  Hou-Jun L{\"u} for providing the data of $ E_{\rm peak} $ of short GRBs, Bin-Bin Zhang for technical help on data processing, Junchen Wan for communications on statistics, George Rhee and Kayhan Gultekin for discussions on black hole number density redshift distribution, and Amy Lien for helpful communications. This work is partially supported by National Basic Research Program (973 Program) of China under Grant No. 2014CB845800, the NSFC (11273005) and SRFDP (20120001110064). Hui Sun is supported by China Scholarship Program to conduct research at UNLV.

\clearpage

\begin{table}
\caption{Instrumental parameters.\label{tab:instruments}}
\begin{center}
\begin{tabular}{llll}
\hline
\hline 
Detectors  & Operation Time & Field of View & Sensitivity ($\rm erg\,cm^{-2}\,s^{-1}$)\\ 
(Instrument)&(T)&($\Omega$)&($ F_{\rm th} $)\\
\hline 
CGRO(BATSE) & 10 yrs & $\pi$ sr. & $3.0\times 10^{-8}$ \\ 
HETE-II(WXM) & 7 yrs & 0.8 sr. & $8.0\times 10^{-9}$ \\ 
INTEGRAL(IBIS) & 12 yrs & 0.26 sr. & $9.1\times10^{-9}$ \\  
Swift(XRT) & 10 yrs &$5\times 10^{-5} sr.$ & $ 10^{-12} $ ($\rm 1000s$)\\
ROSAT(PSPC) & 8 yrs & $10^{-3}$ sr. & $3.0\times 10^{-13}$ ($\rm 500s$) \\ 
XMM-Newton(EPIC) & 15 yrs & $2\times 10^{-4}$ sr. & $2.0\times 10^{-14}$ ($\rm 10^3s$) \\ 
Chandra(ACIS) & 14 yrs & $6\times 10^{-4}$ sr. & $4.0\times 10^{-15}$ ($\rm 10^5s$) \\ 
\hline 

Swift(BAT) & 10 yrs & 1.33 sr. & $ 3\times 10^{-8} $ for HL-LGRBs and rate-triggered LL-GRBs\\
Swift(BAT) & 10 yrs & 1.33 sr. & $  10^{-7} $ for SGRBs\\
Swift(BAT) & 10 yrs & 1.33 sr. & $ 2.8(3.1)\times 10^{-9} $ for LL-LGRB 060218/100316D\\
Swift(BAT) & 10 yrs & 1.33 sr. & $ 10^{-8} $ for Sw J1644+57\\
Swift(BAT) & 10 yrs & 1.33 sr. & $ 4.3\times 10^{-11} $ for Sw J2058+05\\
\hline 
\end{tabular} 
\end{center}
\end{table}

\begin{table}
\caption{Best-fit merger delay models of SGRBs with respect to star formation history.}\label{tab:DL} 
\begin{tabular}{l|c|c|c}
\hline
\hline
Delay Model & Formula & Best-fit parameters & Reference\\
\hline
Gaussian (G)& $ m_{G}(\tau)d\tau=\exp\left( -\frac{(\tau-t_{\rm d,G})^{2}}{{2\sigma_{\rm t,G}^{2}}}\right) /{\sqrt{2\pi}\sigma_{\rm t,G}}d\tau $ & $t_{\rm t,G}=2~{\rm Gyr}$, $\sigma_{\rm t,G}=0.3 $ &(1)\\
Log-normal (LN)& $ m_{\rm LN}(\tau)d\ln\tau=\exp\left( -\frac{(\ln\tau-\ln t_{\rm d,LN})^{2}}{{2\sigma_{\rm t,LN}^{2}}}\right) /({\sqrt{2\pi}\sigma_{\rm t,LN}})d\ln\tau $& $t_{\rm t,LN}=2.9~{\rm Gyr}$, $\sigma_{\rm t,LN}=0.2 $ & (2)\\
Power law (PL) &$ g_{\rm PL}(\tau)d\tau=\tau^{-\alpha_{t}}d\tau $&$\alpha_{t}$=0.81 &(2)\\
\hline
\end{tabular}
\tablerefs{(1).\cite{Vir11}; (2).\cite{Wan14}.}
\end{table}

\begin{table}
\caption{The LL-LGRB sample.\label{tab:GRB}}
\begin{center}
\begin{tabular}{cccccccc}
\hline 
\hline
Name & Detector & Energy band & $P_{\rm peak}$ \tablenotemark{a} & $L_{p,\rm bol,48}$ \tablenotemark{b} & redshift & $E_{\rm peak}$ \tablenotemark{c} & Reference\tablenotemark{d} \\ 
\hline 
GRB 980425 & CGRO & 50-300 keV & $0.96\pm 0.05$ & 0.058 & 0.0085 & 122 keV (CPL) & (1)\\ 

XRF 020903 & HETE II & 2-10 keV & $2.2\pm 0.8$ & 7.42 & 0.251  & 2.6 keV (CPL) & (2) \\ 

GRB 031203 & INTEGRAL & 20-200 keV & $1.3\pm 0.0$ & 9.85 & 0.155  & 121 keV (PL) & (1) \\ 
 
GRB 060218 & Swift & 15-150 keV & $ 0.25\pm 0.11 $ & 0.147 & 0.033 & 4.5 keV (CPL) & (3) \\ 

GRB 080517 & Swift & 15-150 keV & $ 0.6 \pm 0.2 $ & 3.03 & 0.09 & 202 keV (PL) & (4) \\ 

GRB 100316D & Swift & 15-150 keV & $ 0.1\pm 0.0 $ & 0.116 & 0.0591 & 19.6 keV (CPL) & (5) \\ 
\hline 
\end{tabular} 
\tablenotetext{a}{Peak photon flux in unit of $\rm ph\,cm^{-1}\,s^{-1}$. The values for GRB 980425 and GRB 031203 are taken from GCN 67 and GCN 2460 seperately. XRF 020903 is from the \cite{Sak04}. Swift samples are downloaded from swift table ($http://swift.gsfc.nasa.gov/archive/grb\_table/$).}
\tablenotetext{b}{Peak bolometric luminosity is calculated after k-correction. $L_{p,\rm bol,48}$ is in unit of $\rm 10^{48}\,erg\,s^{-1}$.}
\tablenotetext{c}{$E_{\rm peak}$ is either directly given from spectrum fit in literatures for Power-law with a cutoff (CPL)-fit models or calculated through \cite{Vir12} for Power-law (PL)-fit models. }
\tablerefs{(1).\cite{Kan07}; (2).\cite{Sak04}; (3).\cite{Cam06}; (4).\cite{Sta14}; (5).\cite{Fan11}}
\end{center}
\end{table}

\begin{table}
\caption{The shock breakout sample.\label{tab:SBO}}
\begin{center}
\begin{tabular}{lccccc}
\hline 
\hline
Name & Detector & Energy band & $L_{p,\rm bol,46}$ \tablenotemark{a} & redshift & Reference \\ 
\hline 
SN 2006aj/GRB 060218 & Swift (BAT) & 15-150 keV & 14.7 & 0.033 & \cite{Cam06} \\ 
SN 2008D/XRO 080109 & Swift (XRT) & 2-10 keV & 0.0061 & 0.007 & \cite{Sod08} \\ 
\hline
\end{tabular} 
\tablenotetext{a}{The bolometric luminosity of XRO 080109 is calculated from $k$-correction based on a power-law spectral with photon index of 2.3. $L_{p,\rm bol,46}$ is in unit of $\rm 10^{46}\,erg\,s^{-1}$} 
\end{center}
\end{table}

\begin{table}
\caption{The TDE sample.\label{tab:TDE}}
\begin{center}
\begin{tabular}{lcccll}
\hline
\hline 
Name & Detector & Energy band & $L_{p,\rm bol}$\tablenotemark{a} & redshift & Reference\\
\hline
NGC 5905 & ROSAT & 0.1-2.4 keV & $ 3.6\times10^{43} $ & 0.011 & \cite{Bad96}; \cite{KB99} \\
RX J1242& ROSAT & 0.1-2.4 keV & $ 1.2\times10^{44} $ & 0.05 & \cite{KG99} \\
RX J1624 & ROSAT & 0.1-2.4 keV & $ 2.2\times10^{44} $ & 0.064 & \cite{Gru99}\\
RX J1420 & ROSAT & 0.1-2.4 keV & $ 7.6\times10^{43} $ & 0.147 & \cite{Gre00} \\
NGC 3599 & XMM& 0.2-2.0 keV & $ 7.1\times10^{41} $ & 0.0028 & \cite{Esq07} \cite{Esq08} \\
SDSS J1324 & XMM& 0.2-2.0 keV & $ 6.7\times10^{43} $ & 0.088 & \cite{Esq07} \cite{Esq08} \\
TDXFJ 1347&ROSAT & 0.3-2.4 keV & $8.8\times10^{42} $ & 0.037 & \cite{Cap09}\\
SDSSJ 1311&Chandra& 0.3-3.0 keV &$7.0\times10^{42}$ & 0.195 & \cite{Mak10}\\
2XMMiJ 1847 & XMM &0.2-2.0 keV &$3.9\times10^{43}$ & 0.035 &\cite{Lin11} \\
SDSSJ 1201&XMM & 0.2-2.0 keV &$4.2\times10^{44}$ & 0.146 &\cite{Sax12} \\
WINGSJ 1348 & Chandra & 0.2-2.0 keV & $2.8\times10^{42}$ & 0.062 & \cite{Mak13}\\
Swift J1644+57&Swift& 15-150 keV&$7.2\times10^{48}$&0.354&\cite{Blo11}; \cite{Bur11}\\
Swift J2058+05&Swift& 15-150 keV&$7.6\times10^{47}$&1.185&\cite{Cen12}\\
\hline
\end{tabular} 
\tablenotetext{a}{$L_{p,\rm bol}$ is in unit of $\rm erg\,s^{-1} $.}
\end{center}
\end{table}

\renewcommand\arraystretch{1.5}
\begin{table}
\caption{The best fit luminoity functions of different types of extra-galactic high energy transients. For SGRBs, the results for three merger delay models (Gaussian (G), log-normal (LN) and power law (PL)) are given. The 1$\sigma$ errors of all the fitting parameters are presented based on 5000 sets of Monte Carlo simulations. \label{tab:LF}}
\begin{center}
\begin{tabular}{c|c|ccc|c}
\hline 
\hline
Type  & Fit model & $ \alpha_1 $ & $ \alpha_2 $ & $ \alpha_3 $ & $ L_b $ $ (\rm erg\,s^{-1})  $\\ 
\hline
HL-LGRBs & TPL & $2.2^{+0.4}_{-0.2}$ &$1.0^{+0.1}_{-0.1}$ & $2.0^{+0.3}_{-0.3}$ & $ 5.0^{+3.0}_{-1.3}\times 10^{50} $, $ 7.1^{+4.3}_{-3.0}\times 10^{52} $\\ 
LL-LGRBs  & SPL & $2.3^{+0.2}_{-0.2}$ & -&- & - \\ 
Joint HL-/LL-LGRBs & TPL & $1.7^{+0.1}_{-0.1}$&$1.0^{+0.2}_{-0.1}$ & $2.0^{+0.3}_{-0.3}$& $ 1.0^{+0.2}_{-0.3}\times 10^{51} $, $ 7.8^{+2.3}_{-3.1}\times 10^{52} $ \\
SGRBs(G) & SPL & $1.7^{+0.08}_{-0.08}$ & - & - & -\\ 
SGRBs(LN) &  SPL & $1.6^{+0.08}_{-0.08}$ & - & - & -\\ 
SGRBs(PL) &  SPL & $1.5^{+0.08}_{-0.08}$ & - & - & -\\ 
Joint SBO/LL-LGRB & SPL &  $2.0^{+0.09}_{-0.09}$ & - & -  & - \\
TDEs  & SPL & $2.0^{+0.05}_{-0.05}$ & - & -  & -\\  
\hline 
\end{tabular} 
\end{center}
\end{table}

\renewcommand\arraystretch{1.5}
\begin{table}
\caption{The event rate density of various transients given an observed minimum luminosity threshold and a typical luminosity threshold. \label{tab:rho}}
\begin{center}
\begin{tabular}{c|c|c|c|c}
\hline 
\hline
Type & $ L_{m}$ $\rm (erg\,s^{-1})$ & $ \rho_{0,>L_{m}}$ $(\rm Gpc^{-3}yr^{-1})$  & $ L^{'} $ $ (\rm erg\,s^{-1})  $ & $ \rho_{0,>L^{'}}$ $(\rm Gpc^{-3}yr^{-1})$\\ 
\hline
HL-LGRBs  & $ 3\times 10^{49}$ & $ 2.4^{+0.3}_{-0.3} $ & $ 10^{50} $ & $ 0.8^{+0.1}_{-0.1} $  \\
LL-LGRBs  & $ 5\times 10^{46}$ & $  164^{+98}_{-65}$ & $  10^{46} $ & $  440^{+264}_{-175}$ \\  
SGRBs(G)& $ 7\times 10^{49}$ & $ 4.2^{+1.3}_{-1.0} $  & $ 10^{50} $ & $ 1.3^{+0.4}_{-0.3}  $\\ 
SGRBs(LN)& $7\times 10^{49}$  & $ 3.9^{+1.2}_{-0.9} $  & $ 10^{50} $  & $ 1.2^{+0.4}_{-0.3} $\\ 
SGRBs(PL)& $ 7\times 10^{49}$ & $ 7.1^{+2.2}_{-1.7} $  & $ 10^{50} $ & $  3.3^{+1.0}_{-0.8} $\\ 
SBOs & $ 10^{44}$ \tablenotemark{a} & $ 1.9^{+2.4}_{-1.2} \times 10^{4} $  & $ 10^{47} $  & $ 14^{+32}_{-11}$  \\ 
TDEs & $10^{42}$  & $1.0^{+0.4}_{-0.3} \times 10^{5}$  & $ 10^{44}$ & $ 4.8^{+3.2}_{-2.1}\times10^2 $ \\ 
Swift TDEs & $10^{48}$ & $0.03^{+0.04}_{-0.02} $ &  -  &-\\ 
\hline 
\end{tabular} 
\tablenotetext{a}{Since there are only two confirmed SBOs, two characteristic luminosities are given around the exact luminosities of the two events.}
\end{center}
\end{table}

\begin{table}
\caption{The best fit Luminosity function parameters in different redshift bins for full-sample and sub-sample ($P_p>1.8~\rm~ph~cm^{-2}~s^{-1}$).}
\begin{center}

\begin{tabular}{c|c|c|c|c|c}
\hline 
\multicolumn{5}{c}{Full-Sample} \\ 
\hline 
Redshift & $\alpha_1$ & $\alpha_2$ & $L_b$ ($\rm erg~s^{-1}$) & $\rm \rho^z_{0,>L_m}$ ($\rm Gpc^{-3}~yr^{-1}$)\tablenotemark{a}&  $ L_m$ ($\rm erg~s^{-1}$) \\ 
\hline 
$0<z<1$ & 1.6 & 1.8 & $4.0\times 10^{51}$ & $1.6^{+0.3}_{-0.2} $ ($0.3^{+0.1}_{-0.1}$)& $ 10^{50}$ ($ 10^{51}$) \\ 
$0.5<z<1.5$ & 1.4 &1.7 &$ \geqslant 2.0\times 10^{53}$, or $\leqslant 5.1\times 10^{49}$ & $1.2^{+0.2}_{-0.2} $ ($0.4^{+0.1}_{-0.1}$)& $ 10^{50}$ ($ 10^{51}$) \\ 
$1<z<3$ &1.1 & 1.6 & $1.2\times 10^{52}$ & $0.4^{+0.1}_{-0.1} $ & $ 10^{51}$ \\ 
$2<z<4$ & 1.5 & $ >4.0 $ & $ \geqslant 5.5\times 10^{53}$, or $\leqslant 2.0\times 10^{51}$ &  $0.3^{+0.1}_{-0.1} $ & $ 10^{52}$ \\ 
$z>3$ & $\sim0$ & 1.7 & $\leqslant 1.7\times 10^{52}$ &  $0.2^{+0.1}_{-0.1} $ & $ 10^{53}$ \\ 
\hline 
\multicolumn{5}{c}{Sub-Sample} \\ 
\hline 
Redshift & $\alpha_1$ & $\alpha_2$ & $L_b$ ($\rm erg~s^{-1}$) & $\rm \rho^z_{0,>L_m}$ ($\rm Gpc^{-3}~yr^{-1}$)& $ L_m$ ($\rm erg~s^{-1}$) \\ 
\hline 
$0<z<1$ & 0.1\tablenotemark{b} & 1.8 &  $8.5\times 10^{50}$ & $ 1.7^{+0.3}_{-0.2} $ ($ 0.7^{+0.1}_{-0.1} $)& $ 10^{50}$ ($ 10^{51}$)\\ 
$0.5<z<1.5$ & 1.5 & - & $ \geqslant 10^{53}$, or $\leqslant 8.0\times 10^{50}$ & $ 0.7^{+0.1}_{-0.1} $& $ 10^{51}$ \\ 
$1<z<3$ & 1.1 & 1.6 & $3.2\times 10^{52}$ & $ 0.6^{+0.1}_{-0.1}$ & $ 10^{51}$\\ 
$2<z<4$ & 1.0 & 1.5 & $4.3\times 10^{52}$ & $ 0.6^{+0.1}_{-0.1}$ & $ 10^{52}$  \\ 
$z>3$ & $\sim0$ & 1.9 & $7.0\times 10^{52}$ & $ 0.4^{+0.1}_{-0.1}$ & $ 10^{53}$ \\ 
\hline 
\end{tabular} 
\end{center}
\tablenotetext{a} {The local event rate density derived from the data in different redshift bins. The minimum luminosity varies at each bin due to the limited instrument sensitivity. We also give event rate density at same $L_m$ for comparison.}
\tablenotetext{b} {In the redshift bin $0<z<1$, LF could be fit by a TPL. Here $\alpha_1$ and $\alpha_2$ denote the latter two components and $L_b$ is the second break luminosity.}
\label{tab:LFApp}
\end{table}

\clearpage

\clearpage

\begin{thebibliography}{}
\bibitem[Bade et al.(1996)]{Bad96} Bade, N., Komossa, S., \& Dahlem, M.\ 1996, \aap, 309, L35
\bibitem[Band et al.(1993)]{Ban93} Band, D., Matteson, J., 
Ford, L., et al.\ 1993, \apj, 413, 281 
\bibitem[Barniol Duran et al.(2014)]{BD14} Barniol Duran, 
R., Nakar, E., Piran, T., \& Sari, R.\ 2014, arXiv:1407.4475
\bibitem[Barthelmy et al.(2005)]{barthelmy05} Barthelmy, S.~D., 
Chincarini, G., Burrows, D.~N., et al.\ 2005, \nat, 438, 994
\bibitem[Berger(2014)]{berger14} Berger, E.\ 2014, \araa, 52, 43 
\bibitem[Bersier et al.(2006)]{Ber06} Bersier, D., Fruchter, 
A.~S., Strolger, L.-G., et al.\ 2006, \apj, 643, 284
\bibitem[Bloom et al.(2011)]{Blo11} Bloom, J.~S., Giannios, 
D., Metzger, B.~D., et al.\ 2011, Science, 333, 203
\bibitem[Bromberg et al.(2011)]{Bro11} Bromberg, O., Nakar, 
E., \& Piran, T.\ 2011, \apjl, 739, LL55
\bibitem[Bromberg et al.(2013)]{bromberg13} Bromberg, O., Nakar, 
E., Piran, T., \& Sari, R.\ 2013, \apj, 764, 179 
\bibitem[Burrows et al.(2011)]{Bur11} Burrows, D.~N., Kennea, 
J.~A., Ghisellini, G., et al.\ 2011, \nat, 476, 421
\bibitem[Butler et al.(2007)]{Butler07} Butler, N.~R., Kocevski, 
D., Bloom, J.~S., \& Curtis, J.~L.\ 2007, \apj, 671, 656 
\bibitem[Campana et al.(2006)]{Cam06} Campana, S., Mangano, 
V., Blustin, A.~J., et al.\ 2006, \nat, 442, 1008 
\bibitem[Cappelluti et al.(2009)]{Cap09} Cappelluti, N., Ajello, M., Rebusco, P., et al.\ 2009, \aap, 495, L9 
\bibitem[Cenko et al.(2012)]{Cen12} Cenko, S.~B., Krimm, 
H.~A., Horesh, A., et al.\ 2012, \apj, 753, 77 
\bibitem[Colgate(1975)]{Col75} Colgate, S.~A.\ 1975, Seventh 
Texas Symposium on Relativistic Astrophysics, 262, 34 
\bibitem[Donnarumma \& Rossi(2015)]{Donnarumma15} Donnarumma, I. \& Rossi, E. M. 2015, arXiv:1501.05111
\bibitem[Efron 
\& Petrosian(1992)]{EP92} Efron, B., \& Petrosian, V.\ 1992, \apj, 399, 345
\bibitem[Esquej et al.(2007)]{Esq07} Esquej, P., Saxton, R.~D., Freyberg, M.~J., et al.\ 2007, \aap, 462, L49
\bibitem[Esquej et al.(2008)]{Esq08} Esquej, P., Saxton, R.~D., Komossa, S., et al.\ 2008, \aap, 489, 543 
\bibitem[Fan et al.(2011)]{Fan11} Fan, Y.-Z., Zhang, B.-B., 
Xu, D., Liang, E.-W., \& Zhang, B.\ 2011, \apj, 726, 32
\bibitem[Fong et al.(2010)]{fong10} Fong, W., Berger, E., 
\& Fox, D.~B.\ 2010, \apj, 708, 9
\bibitem[Fong 
\& Berger(2013)]{fong13} Fong, W., \& Berger, E.\ 2013, \apj, 776, 18
\bibitem[Fox et al.(2005)]{fox05} Fox, D.~B., Frail, D.~A., 
Price, P.~A., et al.\ 2005, \nat, 437, 845
\bibitem[Gehrels(1986)]{Geh86} Gehrels, N.\ 1986, \apj, 303, 
336 
\bibitem[Gehrels et al.(2004)]{gehrels04} Gehrels, N., 
Chincarini, G., Giommi, P., et al.\ 2004, \apj, 611, 1005 
\bibitem[Greiner et al.(2000)]{Gre00} Greiner, J., Schwarz, R., Zharikov, S., \& Orio, M.\ 2000, \aap, 362, L25

\bibitem[Grupe et al.(1999)]{Gru99} Grupe, D., Thomas, H.-C., \& Leighly, K.~M.\ 1999, \aap, 350, L31
\bibitem[Hjorth \& Bloom (2011)]{HB11} Hjorth, J., \& Bloom, J. S. 2011, in Gamma-ray Bursts, ed. C. Kouveliotou,
R. A. M. J. Wijers, \& S. E. Woosley (Cambridge: Cambridge Univ. Press),
chapter 9
\bibitem[Hopkins 
\& Beacom(2006)]{HB06} Hopkins, A.~M., \& Beacom, J.~F.\ 2006, \apj, 651, 142
\bibitem[Kaneko et al.(2007)]{Kan07} Kaneko, Y., Ramirez-Ruiz, E., Granot, J., et al.\ 2007, \apj, 654, 385
\bibitem[Kistler et al.(2008)]{Kistler08} Kistler, M.~D., 
Y{\"u}ksel, H., Beacom, J.~F., \& Stanek, K.~Z.\ 2008, \apjl, 673, L119
\bibitem[Klein \& Chevalier(1978)]{KC78} Klein, R.~I., \& Chevalier, R.~A.\ 1978, \apjl, 223, L109
\bibitem[Kocevski 
\& Liang(2006)]{KL06} Kocevski, D., \& Liang, E.\ 2006, \apj, 642, 371 
\bibitem[Komossa \& Bade(1999)]{KB99} Komossa, S., \& Bade, N.\ 1999, \aap, 343, 775 
\bibitem[Komossa \& Greiner(1999)]{KG99} Komossa, S., \& Greiner, J.\ 1999, \aap, 349, L45 
\bibitem[Komossa(2012)]{Kom12} Komossa, S.\ 2012, European 
Physical Journal Web of Conferences, 39, 02001 
\bibitem[Kouveliotou et al.(1993)]{Kou93} Kouveliotou, C., 
Meegan, C.~A., Fishman, G.~J., et al.\ 1993, \apjl, 413, L101 
\bibitem[Kumar \& Zhang(2015)]{KZ15} Kumar, P., \& Zhang, B.\ 2015, Phys. Rep., 561, 1
\bibitem[Lei \& Zhang(2011)]{LZ11} Lei, W.-H., \& Zhang, B.\ 2011, \apjl, 740, LL27 
\bibitem[Lei et al.(2013)]{Lei13} Lei, W.-H., Zhang, B., \& Gao, H. \ 2013, \apj, 762, 98
\bibitem[Liang et al.(2007)]{Liang07} Liang, E., Zhang, B., 
Virgili, F., \& Dai, Z.~G.\ 2007, \apj, 662, 1111 
\bibitem[Lien et al.(2014)]{Lie14} Lien, A., Sakamoto, T., 
Gehrels, N., et al.\ 2014, \apj, 783, 24 
\bibitem[Li(2008)]{li08} Li, L.-X.\ 2008, \mnras, 388, 1487 
\bibitem[Lin et al.(2011)]{Lin11} Lin, D., Carrasco, E.~R., 
Grupe, D., et al.\ 2011, \apj, 738, 52
\bibitem[Liu et al.(2015)]{Liu15} Liu, D., Pe'er, A., \& Loeb, A. 2015, \apj, 798, 13
\bibitem[Lloyd-Ronning et al.(2002)]{LR02} Lloyd-Ronning, 
N.~M., Fryer, C.~L., \& Ramirez-Ruiz, E.\ 2002, \apj, 574, 554
\bibitem[Lodato et al.(2009)]{Lodato09} Lodato, G., King, A. R., \& Pringle, J. E. 2009, \mnras, 392, 332
\bibitem[Luo et al.(2008)]{Luo08} Luo, B., Brandt, W.~N., 
Steffen, A.~T., \& Bauer, F.~E.\ 2008, \apj, 674, 122 
\bibitem[L{\"u} et al.(2014)]{Lv14} L{\"u}, H.-J., Zhang, 
B., Liang, E.-W., Zhang, B.-B., \& Sakamoto, T.\ 2014, \mnras, 442, 1922
\bibitem[L{\"u} et al.(2015)]{Lv15} L{\"u}, H.-J., Zhang, 
B., Lei, W.-H., Li, Y., \& Lasky, P.~D 2015, \apj, 805, 89
\bibitem[Maksym et al.(2010)]{Mak10} Maksym, W.~P., Ulmer, 
M.~P., \& Eracleous, M.\ 2010, \apj, 722, 1035
\bibitem[Maksym et al.(2013)]{Mak13} Maksym, W.~P., Ulmer, 
M.~P., Eracleous, M.~C., Guennou, L., \& Ho, L.~C.\ 2013, \mnras, 435, 1904
\bibitem[Metzger et al.(2012)]{Metzger12} Metzger, B. D., Giannios, D., \& Mimica, P. 2012, \mnras, 420, 3528
\bibitem[Nakar \& Sari(2010)]{NS10} Nakar, E., \& Sari, R.\ 2010, \apj, 725, 904
\bibitem[Nakar \& Sari(2012)]{NS12} Nakar, E., \& Sari, R.\ 2012, \apj, 747, 88
\bibitem[Paciesas et al.(1999)]{paciesas99} Paciesas, W.~S., 
Meegan, C.~A., Pendleton, G.~N., et al.\ 1999, \apjs, 122, 465
\bibitem[Pescalli et al.(2015)]{Pescalli15} Pescalli, A., 
Ghirlanda, G., Salvaterra, R., et al.\ 2015, arXiv:1506.05463  
\bibitem[Petrosian et al.(2015)]{Petro15} Petrosian, V., 
Kitanidis, E., \& Kocevski, D.\ 2015, \apj, 806, 44
\bibitem[Planck Collaboration et al.(2015)]{Pla15} Planck 
Collaboration, Ade, P.~A.~R., Aghanim, N., et al.\ 2015, arXiv:1502.01589 
\bibitem[Qin et al.(2010)]{Qin10} Qin, S.-F., Liang, E.-W., 
Lu, R.-J., Wei, J.-Y., \& Zhang, S.-N.\ 2010, \mnras, 406, 558 
\bibitem[Qin et al.(2013)]{qin13} Qin, Y., Liang, E.-W., 
Liang, Y.-F., et al.\ 2013, \apj, 763, 15 
\bibitem[Rees(1988)]{Ree88} Rees, M.~J.\ 1988, \nat, 333, 523
\bibitem[Robertson 
\& Ellis(2012)]{Robertson12} Robertson, B.~E., \& Ellis, R.~S.\ 2012, \apj, 744, 95 
\bibitem[Sakamoto et al.(2004)]{Sak04} Sakamoto, T., Lamb, 
D.~Q., Graziani, C., et al.\ 2004, \apj, 602, 875
\bibitem[Sakamoto et al.(2008)]{Sak08} Sakamoto, T., 
Barthelmy, S.~D., Barbier, L., et al.\ 2008, \apjs, 175, 179 
\bibitem[Sakamoto et al.(2009)]{Sak09} Sakamoto, T., Sato, 
G., Barbier, L., et al.\ 2009, \apj, 693, 922  
\bibitem[Sakamoto et al.(2011)]{sakamoto11} Sakamoto, T., 
Barthelmy, S.~D., Baumgartner, W.~H., et al.\ 2011, \apjs, 195, 2 
\bibitem[Salvaterra et al.(2009)]{Sal09} Salvaterra, R., 
Guidorzi, C., Campana, S., Chincarini, G., 
\& Tagliaferri, G.\ 2009, \mnras, 396, 299
\bibitem[Salvaterra et al.(2012)]{Sal12} Salvaterra, R., 
Campana, S., Vergani, S.~D., et al.\ 2012, \apj, 749, 68
\bibitem[Saxton et 
al.(2008)]{Sax08} Saxton, R.~D., Read, A.~M., Esquej, P., et al.\ 2008, \aap, 480, 611
\bibitem[Saxton et 
al.(2012)]{Sax12} Saxton, R.~D., Read, A.~M., Esquej, P., et al.\ 2012, \aap, 541, AA106
\bibitem[Shankar et al.(2013)]{Sha13} Shankar, F., Weinberg, 
D.~H., \& Miralda-Escud{\'e}, J.\ 2013, \mnras, 428, 421 
\bibitem[Soderberg et al.(2006)]{Sod06} Soderberg, A.~M., Kulkarni, S.~R., Nakar, E., et al.\ 2006, \nat, 442, 1014
\bibitem[Soderberg et al.(2008)]{Sod08} Soderberg, A.~M., Berger, E., Page, K.~L., et al.\ 2008, \nat, 453, 469 
\bibitem[Stanway et al.(2014)]{Sta14} Stanway, E.~R., Levan, 
A.~J., Tanvir, N.~R., et al.\ 2014, arXiv:1409.5791 
\bibitem[Tchekhovskoy et al.(2014)]{Tch14} Tchekhovskoy, A., Metzger, B. D., Giannios, D., Kelley, L. Z.
2014, \mnras, 437, 2744
\bibitem[van Velzen et 
al.(2013)]{Van13} van Velzen, S., Frail, D.~A., K{\"o}rding, E., \& Falcke, H.\ 2013, \aap, 552, AA5
\bibitem[Virgili et al.(2009)]{Vir09} Virgili, F.~J., Liang, 
E.-W., \& Zhang, B.\ 2009, \mnras, 392, 91 
\bibitem[Virgili et al.(2011)]{Vir11} Virgili, F.~J., Zhang, 
B., O'Brien, P., \& Troja, E.\ 2011, \apj, 727, 109 
\bibitem[Virgili et al.(2011b)]{Virgili11b} Virgili, F.~J., Zhang, 
B., Nagamine, K., \& Choi, J.-H.\ 2011b, \mnras, 417, 3025
\bibitem[Virgili et al.(2012)]{Vir12} Virgili, F.~J., Qin, 
Y., Zhang, B., \& Liang, E.\ 2012, \mnras, 424, 2821 
\bibitem[Voges et al.(1999)]{Vog99} Voges, W., Aschenbach, B., Boller, T., et al.\ 1999, \aap, 349, 389 
\bibitem[Wanderman \& Piran(2010)]{Wan10} Wanderman, D., \& Piran, T.\ 2010, \mnras, 406, 1944 
\bibitem[Wanderman \& Piran(2014)]{Wan14} Wanderman, D., \& Piran, T.\ 2014, arXiv:1405.5878 
\bibitem[Wang \& Merritt(2004)]{WM04} Wang, J., \& Merritt, D.\ 2004, \apj, 600, 149
\bibitem[Wang et al.(2014)]{Wang14} Wang, J.-Z., Lei, W.-H., Wang, D.-X. et al. 2014, \apj, 788, 32
\bibitem[Wang et al.(2007)]{wangxy07} Wang, X.-Y., Li, Z., 
Waxman, E., \& M{\'e}sz{\'a}ros, P.\ 2007, \apj, 664, 1026 
\bibitem[Waxman et al.(2007)]{Wax07} Waxman, E., 
M{\'e}sz{\'a}ros, P., \& Campana, S.\ 2007, \apj, 667, 351 
\bibitem[Woosley \& Weaver(1986)]{WW86} Woosley, S.~E., \& Weaver, T.~A.\ 1986, \araa, 24, 205 
\bibitem[Yu et al.(2015)]{Yu15} Yu, H., Wang, F.~Y., Dai, 
Z.~G., \& Cheng, K.~S.\ 2015, \apjs, 218, 13
\bibitem[Yuan et al.(2015)]{Yuan15} Yuan, W., Zhang, C., Feng, H., Zhang, S. N., et al. arXiv:1506.07735
\bibitem[Yonetoku et al.(2004)]{Yone04} Yonetoku, D., 
Murakami, T., Nakamura, T., et al.\ 2004, \apj, 609, 935
\bibitem[Y{\"u}ksel et al.(2008)]{Yuk08} Y{\"u}ksel, H., 
Kistler, M.~D., Beacom, J.~F., \& Hopkins, A.~M.\ 2008, \apjl, 683, L5 
\bibitem[Zhang et al.(2004)]{Zha04} Zhang, B., Dai, X., Lloyd-Ronning, N. M., M\'esz\'aros, P. 2004, \apj, 601, L119
Liang, E.-W., et al.\ 2007, \apjl, 655, L25 
\bibitem[Zhang et al.(2007)]{Zhang07} Zhang, B., Zhang, B.-B., 
Liang, E.-W., et al.\ 2007, \apjl, 655, L25 
\bibitem[Zhang et al.(2009)]{Zhang09} Zhang, B., Zhang, B.-B., 
Virgili, F., et al.\ 2009, \apj, 703, 1696 
\bibitem[Zauderer et al.(2011)]{zauderer11} Zauderer, B.~A., 
Berger, E., Soderberg, A.~M., et al.\ 2011, \nat, 476, 425
\end{thebibliography}
\end{document}